\documentclass[twocolumn]{IEEEtran}
\usepackage[latin9]{inputenc}
\usepackage{color}
\usepackage{float}
\usepackage{booktabs}
\usepackage{units}
\usepackage{amsmath}
\usepackage{amsthm}
\usepackage{amssymb}
\usepackage{graphicx}
\usepackage{epstopdf}

\makeatletter

\newcommand{\noun}[1]{\textsc{#1}}
\providecommand{\tabularnewline}{\\}
\floatstyle{ruled}
\newfloat{algorithm}{tbp}{loa}
\providecommand{\algorithmname}{Algorithm}
\floatname{algorithm}{\protect\algorithmname}

\theoremstyle{plain}
\newtheorem{thm}{\protect\theoremname}
\theoremstyle{remark}
\newtheorem{rem}[thm]{\protect\remarkname}

\usepackage{cite}
\usepackage{algorithm}
\usepackage[noend]{algpseudocode}
\usepackage{caption}
\providecommand{\remarkname}{Remark}
\providecommand{\theoremname}{Theorem}

\@ifundefined{showcaptionsetup}{}{%
 \PassOptionsToPackage{caption=false}{subfig}}
\usepackage{subfig}
\makeatother

\providecommand{\remarkname}{Remark}
\providecommand{\theoremname}{Theorem}

\begin{document}
\title{End-to-end Wireless Path Deployment with Intelligent Surfaces Using
Interpretable Neural Networks}
\author{Christos Liaskos\IEEEauthorrefmark{1}\IEEEauthorrefmark{5}, Shuai
Nie\IEEEauthorrefmark{3}, Ageliki Tsioliaridou\IEEEauthorrefmark{1},
Andreas Pitsillides\IEEEauthorrefmark{2}, \\
Sotiris Ioannidis\IEEEauthorrefmark{4}\IEEEauthorrefmark{1}, and
Ian Akyildiz\IEEEauthorrefmark{2}\IEEEauthorrefmark{3}\\
 {\small{}\IEEEauthorrefmark{1}Foundation for Research and Technology
- Hellas (FORTH), emails: \{cliaskos, atsiolia\}@ics.forth.gr}\\
 {\small{}\IEEEauthorrefmark{2}University of Cyprus, Computer Science
Department, email: Andreas.Pitsillides@ucy.ac.cy}\\
{\small{}\IEEEauthorrefmark{3}Georgia Institute of Technology, emails:
\{shuainie, ian\}@ece.gatech.edu}\\
 {\small{}\IEEEauthorrefmark{4}Technical University of Chania, School
of Electrical and Computer Engineering, Greece, email: sotiris@ece.tuc.gr}\\
{\small{}\IEEEauthorrefmark{5}University of Ioannina, Department
of Computer Science and Engineering, Greece.}}
\maketitle
\begin{abstract}
Intelligent surfaces exert deterministic control over the wireless
propagation phenomenon, enabling novel capabilities in performance,
security and wireless power transfer. Such surfaces come in the form
of rectangular tiles that cascade to cover large surfaces such as
walls, ceilings or building facades. Each tile is addressable and
can receive software commands from a controller, manipulating an impinging
electromagnetic wave upon it by customizing its reflection direction,
focus, polarization and phase. A new problem arises concerning the
orchestration of a set of tiles towards serving end-to-end communication
objectives. Towards that end, we propose a novel intelligent surface
networking algorithm based on interpretable neural networks. Tiles
are mapped to neural network nodes and any tile line-of-sight connectivity
is expressed as a neural network link. Tile wave manipulation functionalities
are captured via geometric reflection with virtually rotatable tile
surface norm, thus being able to tunable distribute power impinging
upon a tile over the corresponding neural network links, with the
corresponding power parts acting as the link weights. A feedforward/backpropagate
process optimizes these weights to match ideal propagation outcomes
(normalized network power outputs) to wireless user emissions (normalized
network power inputs). An interpretation process translates these
weights to the corresponding tile wave manipulation functionalities.
\end{abstract}

\begin{IEEEkeywords}
Wireless, Intelligent surfaces, Software control, Neural Network,
Interpretable.
\end{IEEEkeywords}

\section{Introduction\label{sec:intro}}

In recent years, the development of wireless communication technologies
has undergone an unprecedented growth with the proliferation of smart
wireless devices and burgeons of the Internet of Things (IoT). The
goals to achieve higher throughput, lower latency, better services,
and lower prices have driven the research efforts into the fifth and
sixth-generation mobile networks, narrow-band IoT, new generation
WLANs IEEE802.11ax/ay, among other promising techniques to overcome
the challenges of limited spectrum resources and higher user densities~\cite{akyildiz20165g,huang2019survey,ghendir2019towards}.
Most of the currently proposed solutions in the physical layer focus
on the improvement of transceiver design, for instance, the massive
MIMO communication system equips base stations with more than one
hundred antennas to improve the spatial diversity and overall antenna
gain. Nonetheless, techniques on mere transceiver advancements do
not overcome system limitations, and they also suffer from intrinsic
design issues (e.g., self-interference from duplex communications,
channel aging in precoding, and so on). Hence, non-conventional solutions
should be sought.

One important yet largely overlooked factor is the wireless propagation
environment itself, which directly influences the performance of wireless
links. In regular wireless propagation environments, the radiated
electromagnetic (EM) waves experience various effects, including reflection,
diffraction, scattering, penetration, among others, which create rich
multi-path conditions. Such multi-path propagation, if not well-controlled,
could cause destructive interference at the user-ends. Additionally,
energy dissipation occurs with such propagation effects, thus limiting
the received signal strength. In order to minimize such adversarial
effects and preserve energy along transmission, the propagation environments
need to be controlled strategically. With the use of novel materials,
the propagation environments can be transformed into \emph{programmable}
media, yielding unparalleled, wired-level gains in wireless power
transfer, and mitigate interference, Doppler effects, as well as malicious
eavesdropping~\cite{liaskos2018new,liaskos2019network,zhang2019specific}.
In order to perform an adaptive control over the programmable wireless
propagation channel, a novel solution based on artificial materials
has been conceived, utilizing software-defined metasurfaces (SDMs)~\cite{liaskos2015design,oliveri2015reconfigurable}.
\begin{figure}[!tp]
\centering{}\includegraphics[clip,width=1\columnwidth]{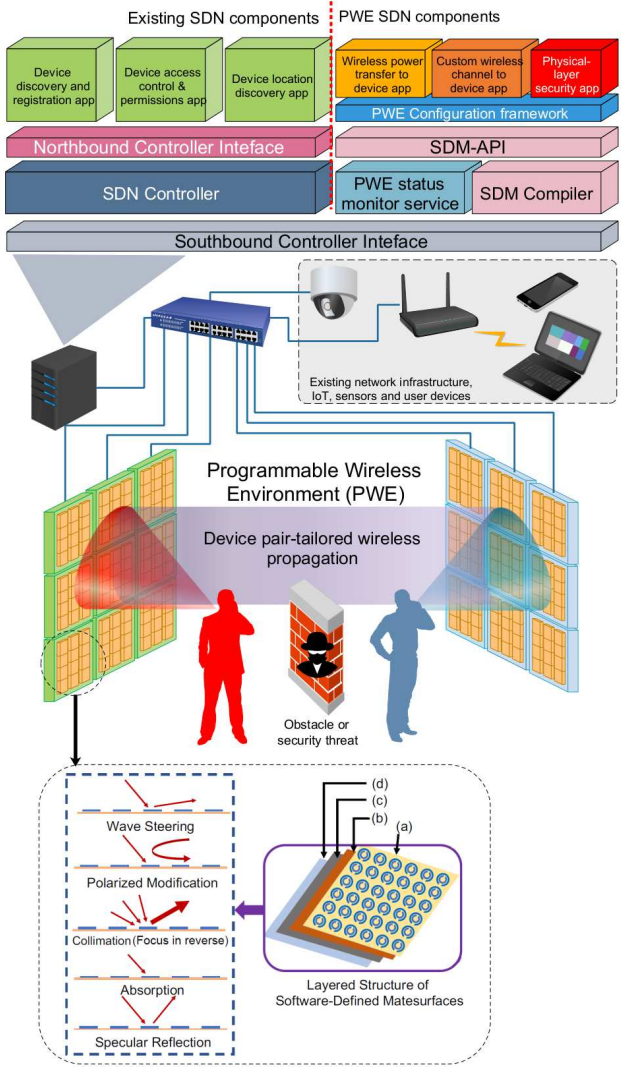}
\caption{\label{fig:SDM}Schematic displaying the inter-networking of SDMs
and their end-to-end workflow, enabling operation modes in a programmable
wireless environment (PWE), illustrated with a pair of a transmitter
and a receiver. The layers of an SDM unit (tile) include (a) the metasurface,
(b) the sensing and actuation, (c) the computing, and (d) the gateway/communications
planes.}
\end{figure}

The SDMs, when embedded within a physical setting, result into a Programmable
Wireless Environment (PWE) wherein wireless propagation can be even
completely customized~\cite{liaskos2019network}. SDMs empower PWEs
by achieving the following functionalities (Fig.~\ref{fig:SDM}):
1) Controlling the propagation direction of the signal. For example,
when EM waves impinge on a flat surface, the law of specular reflection
dictates the direction of reflection. However, the SDMs can steer
the reflected EM waves towards alternate directions, such that useful
signals can be directed towards the destined user, while interference
can be mitigated; 2) Tuning the phase of the signal. Specifically,
during a coherent combining at the receiver side with multi-paths,
signals could be totally canceled out due to a $180^{\circ}$ phase
difference. With the phase-tuning capability of SDMs, the signal level
will be strengthened upon reception; 3) Modifying the polarization
of the signal. While some radiated signals are linearly polarized,
interactions with surrounding environments can alter their polarization.
The polarization mismatch will also cause degradation in link performance.
The SDMs can perform polarization tuning so that the signals are robust
against such distortion.

In this paper, we propose a solution for the networking of sets of
SDM units\textendash denoted as tiles\textendash which are deployed
within an environment, e.g. over indoor walls or building facade.
The problem is to define the exact EM wave manipulation type that
each tile should exert (e.g., steer a wave toward a custom direction),
in order to optimally serve a set of communicating user objectives
(e.g., maximizing the received power). The key idea is that, since
SDM tiles can regulate the distribution of power within a space, they
can be represented by nodes in a neural network, while the wireless
propagation paths can be mapped to neural network links and their
weights. The neural network is optimized via a custom feed-forward/back-propagation
process, and its elements are interpreted into SDM tile functionalities.
The machine learning approach is shown to be intuitive in its representation,
operate collaboratively with existing techniques and be economic in
its use of SDM tiles, compared to related approaches~\cite{liaskos2019network}.\textcolor{black}{{}
The contributions are:}
\begin{itemize}
\item \textcolor{black}{We propose a novel approach that constitutes common
feedforward/backpropagation-based neural networks applicable to the
optimization of PWEs. Using the proposed approach, the structure and
training state of neural networks becomes directly interpretable to
the floorplan geometry and the configuration of its contained SDM
tiles.}
\item \textcolor{black}{Subsequently, we propose a specific scheme for optimizing
PWEs, named }\textcolor{black}{\noun{NNConfig}}\textcolor{black}{;
We map the trained neural networks to tile configurations in order
to achieve EM wave control functionalities, including wave steering,
absorption, and wave splitting. }
\item \textcolor{black}{We demonstrate through extensive simulations that
such interpretable neural networks can deduce PWE configurations that
surpass existing approaches, by detecting complex solutions where
a single tile serves multiple purposes. Subsequently, the number of
tiles required to serve a pair of wireless devices is reduced, benefiting
PWE scalability, capacity and cost. }
\end{itemize}
The remainder of this is work is as follows. Section~\ref{sec:Related-work}
surveys the related studies and provide the necessary prerequisites
on networking intelligent surfaces. Section~\ref{sec:analysis} presents
the novel scheme and its evaluation via simulations takes place in
Section~\ref{sec:Evaluation}. The paper is concluded in Section~\ref{sec:Conclusion}.

\section{Related Work\label{sec:Related-work}}

\begin{figure}[!tp]
\centering{}\includegraphics[clip,width=1\columnwidth]{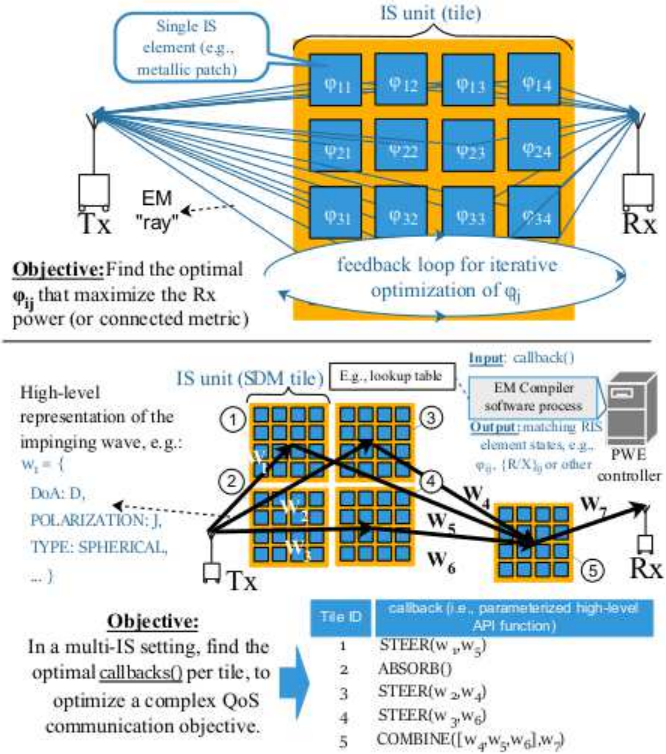}
\caption{\label{fig:SEPCONCERN}\textcolor{black}{Separation of concerns in
the literature. (Top): deducing the optimal meta-atom phases (and
sometime amplitudes) to yield a high level intelligent surface functionality
(e.g., steering). (Bottom): In a multi-surface deployment, deduce
the optimal parameterized high-level functionalities per surface tile,
to serve the user communication objectives.}}
\end{figure}

\textcolor{black}{Presently, the related literature moves into two
disjoint directions, summarized in Fig.~\ref{fig:SEPCONCERN}:}

\textcolor{black}{We will denote the first direction, shown at the
top, as the Intelligent Surface (IS) Physical-Layer. This direction
commonly assumes a single IS and a set of RX-TX. The IS is configured
via the phased antenna array analytical model, meaning that each tile
is treated as an antenna array with individually controlled re-radiation
phase~$\phi_{ij}$ (and in some case, re-radiation amplitude $a_{ij}$
as well)~\cite{huang2019reconfigurable}. The objective of this line
of work is to deduce the optimal~$\phi_{ij}$ values that optimize
the received EM wave, e.g., its power in the simplest case, and extending
to channel quality control in general~\cite{tan2016increasing,wu2018intelligent,hu2017beyond},
(albeit bearing different names, such as ``large intelligent surfaces'',
``intelligent reconfigurable surfaces'', ``holographic MIMO surfaces'',
among others~\cite{zhang2018space,huang2019holographic,han2019intelligent,zhao2019optimizations,cayamcela2018artificially,akyildiz2018combating,nie2019intelligent,liu2019intelligent,huang2018energy}).
This is accomplished by establishing a feedback loop between the RX
and the TX, and applying an iterative optimization process, including
innovative neural networks accelerators that speed up the optimization
process~\cite{huang2019indoor,huang2020reconfigurable}. Other studies
close to this direction seek to optimize the geometry and material
composition of the IS for optimal operation in a given band or application
setting~\cite{samadi2019reconfigurable} (e.g., WiFi). We remark
that a vast number of designs, comprising diverse materials and manufacturing
approaches, has been proposed in the literature over the past 20 years~\cite{tasolamprou2019exploration,liu2019intelligent},
and the reader is directed to a recent survey on this topic~\cite{tsilipakostoward}. }

\textcolor{black}{The second direction, shown at the lower part of
Fig.~\ref{fig:SEPCONCERN}, will be denoted as IS Network-Layer.
This direction assumes a system architecture based on the SDM paradigm,
i.e., networked ISs which are centrally controlled by a server (PWE
controller)~\cite{liaskos2019network,liaskos2018realizing,Liaskos2019ADHOC,liaskos2018new,SPAWC.2019}.
Multiple SDM units are placed within an environment, with the utter
goal of ideally exerting complete control over the wireless propagation
phenomenon within a space (i.e., a PWE). SDMs decouple the functionality-level
multi-IS orchestration logic from the underlying IS Physical-layer
as follows:}
\begin{itemize}
\item \textcolor{black}{The EM waves are modeled at a high-level as data
structures, with an example shown in Fig.~\ref{fig:SEPCONCERN}.
Similarly, the possible/supported EM wave manipulations by a SDM are
modeled as software callbacks, as shown in both Fig.~\ref{fig:SDM}
and Fig.~\ref{fig:SEPCONCERN}. These data structures and callbacks
are collectively denoted as the SDM Application programming Interface
(API)~~\cite{LiaskosAPI}. }
\item \textcolor{black}{A software middleware residing inside the PWE controller,
denoted as the EM Compiler~~\cite{LiaskosComp}, translates callbacks
to matching meta-atom states per tile, such as phases~$\phi_{ij}$,
or even directly to local surface impedance values R/X, when a specific
IS Physical-Layer design/material is considered~\cite{pitilakis2020multi}.
The works in the IS Physical-Layer direction could fill the role of
the EM Compiler. However, to support real-time operation, SDMs use
a lookup table containing all supported functionalities. This table
is populated during the SDM manufacturing, using real field measurements.
A report on the experimental verification of this process can be found
online~\cite{EUrep}. A mechanism for interleaving multiple high-level
functionalities (callbacks) over the same tile on-the-fly is also
supported. In this manner, real-time operation is facilitated, while
the PWE control logic as a whole is decoupled from the multitude of
RIS physical design variations, thus becoming a reusable control module. }
\end{itemize}
\textcolor{black}{Based on this categorization, the present work falls
into the IS Network-Layer direction, and its objective is to define
the optimal high-level functionality (or callback) for each tile within
a PWE, to meet a QoS objective for the communicating user pairs, as
exemplary shown in Fig.~\ref{fig:SDM}-\ref{fig:SEPCONCERN}.}

\textcolor{black}{Based on the discussion above, it is clarified that
the IS Physical-Layer and IS Network-Layer literature directions have
different objectives, which make them complimentary in their nature,
and not otherwisely comparable. }

\subsection{Prerequisites: Networking Metasurfaces\label{subsec:Prerequisites:-Networking-Metasu}}

\textcolor{black}{As shown in Fig.~\ref{fig:SDM}, via their gateways,
the SDMs within a PWE are networked~\cite{Liaskos2018using}, i.e.,
become centrally monitored and configured via a server, in order to
serve a particular end-objective. Specifically, a set of SDMs is configured
with appropriate EM wave steering and focusing commands to route EM
waves exchanged between a pair of wireless users in an unnatural manner,
avoiding obstacles or eavesdroppers~\cite{liaskos2019network}. Other
examples include wireless power transfer and wireless channel customization
for advanced QoS~\cite{ozdogan2019intelligent,Liaskos2018using}.
Using a software-defined networking (SDN)-compatible architecture~\cite{oSDNcontrol},
a PWE can inter-operate with existing SDN applications~\cite{liaskos2019network}.
For instance, a device localization application can inform the PWE
controller of the approximate user device locations within a space~\cite{lemic2019location,lemic2019regression,ABSENSE}.
A user access SDN application can further deduce the access level
that each device should have. Subsequently, the PWE can tune the SDM
tiles to avoid, e.g., potentially malevolent users.}

\textcolor{black}{PWEs have the potential to provide full, software-defined
control over the wireless propagation phenomenon. This can yield groundbreaking
capabilities such as~\cite{zhang2018space,huang2019holographic,han2019intelligent,zhao2019optimizations,cayamcela2018artificially,akyildiz2018combating,nie2019intelligent,liu2019intelligent,huang2018energy,liaskos2019network,Liaskos2019ADHOC}:
i) Extremely efficient mitigation of path loss and multi-path fading
phenomena, allowing not only for more efficient communications, but
also for long-range wireless power transfer. ii) Cross-device interference
cancellation and wireless channel capacity maximization. iii) Physical-layer
security via: SINR minimization around unauthorized users; Improbable
air-route establishment circumventing the location of potential eavesdroppers;
Deliberate accentuation of destructive signal effects (fading, coding
efficiency), localized only in the vicinity of potential eavesdroppers.
iv) Environmental encoding/re-encoding of traveling waves, allowing
for distributed signal processing at unprecedented levels~\cite{zhang2018space}.
v) Highly efficient user device localization, especially in tandem
with 3rd party systems~\cite{ABSENSE}. The full capability of PWEs
is realized in full-scale tile deployments, which attracts attention
in studies exploring the upper bounds of PWE performance~\cite{liaskos2019network}.
However, partial deployments can still yield significant performance
per capability type~\cite{zhang2018space,huang2019holographic,han2019intelligent,akyildiz2018combating,nie2019intelligent,liu2019intelligent,huang2018energy,liaskos2019network,Liaskos2019ADHOC}.
}\textcolor{black}{\emph{Nonetheless, it must be noted that PWEs are
not a replacement for regular communications: not all environments
need to be aware of the wireless propagation and interact with it}}\textcolor{black}{.
For instance, the average home user may not have the economic incentive/practical
benefit to invest in a PWE deployment, since providing plain web access
is usually enough to characterize a home network as satisfactory or
even ideal. This is in contrast to industrial and military settings,
where high capacity, security, localization and mobility requirements
could make large-scale PWEs an appealing option. }

\textcolor{black}{With regard to the authors previous work,~\cite{liaskos2018new}
presented the PWE concept for the first time, but offered no solution
to the PWE configuration problem (i.e., which wave manipulation type
to use per tile to serve a set of users). Similarly,~\cite{Liaskos2019ADHOC}
explored the concept of PWE-enabled security for the first time and
outlined the involved challenges, without defining an algorithm of
tuning a PWE accordingly. A} networking algorithm, denoted as \noun{KpConfig},
was presented by the authors in~\cite{liaskos2019network}. First,
\noun{KpConfig} enforces a graph-based modeling of a PWE as follows.
Each tile is modeled as a graph vertex, and every tile pair that has
line-of-sight connectivity is modeled as a graph edge. Any user device
is also modeled as a graph node, with links connecting it to the \noun{K}
tile-nodes that are affected by its wireless emissions.

Using this graph model of the PWE as a basis, \noun{KpConfig} connects
communicating user devices by finding K-paths within the graph. Every
tile-node along the found paths are then configured with the corresponding
EM wave steering API callback. \noun{KpConfig }offers versatility
in serving many types of user objectives: graph links that are deemed
too close to potential eavesdroppers can be filtered out during the
K-paths finding process; when a user is moving across a trajectory
and is subject to the Doppler effect, the K-path finding process can
consider only his graph-links that are most perpendicular to the trajectory;
phase alteration API callbacks can be applied at a tile in the middle
of a path, thereby performing equalization across all K-paths, etc~\cite{liaskos2019network}.
However, a disadvantage of \noun{KpConfig} is that it applies callbacks
to tiles sequentially. This means that any tile configured for a given
path cannot be re-used to potentially serve many paths at once. Thus,
\noun{KpConfig} can quickly use all available tiles, thereby yielding
limited networking capacity for the PWE.

In differentiation, this study proposes a neural network-based approach
in configuring a PWE, which does not pose a serial configuration restriction.
Moreover, the proposed scheme can work in tandem with \noun{KpConfig},
inheriting its objective-meeting versatility described above (QoS,
Doppler effect mitigation, security versus eavesdroppers). In the
following, we will assume that \noun{KpConfig} has executed (with
any applicable link-filtering taken into account) and a series of
K-paths has been exported for a given user pair. From this information,
we will only keep the tiles found across these paths (which will be
re-organized in layers), while any \noun{KpConfig}-derived tile configuration
(such as steering) will be discarded and overridden by the new scheme.

\textcolor{black}{Moreover, in their previous work of~\cite{SPAWC.2019},
the authors presented a 2D precursor of the present work, which employed
a neural network approach for configuring PWEs, hinting a potentially
economic use of tiles compared to }\textcolor{black}{\noun{KpConfig}}\textcolor{black}{.
In this work we extend~\cite{SPAWC.2019} to operate in full 3D settings,
perform a more complete study of the neural network applicability
to the PWE configuration problem, and perform a thorough comparison
to the }\textcolor{black}{\noun{KpConfig}}\textcolor{black}{{} algorithm
which, to the best of our knowledge is the only related work for the
network-layer PWE configuration.}

\section{Configuration of PWEs Using Artificial Neural Networks\label{sec:analysis}}

In this section, we describe the artificial neural network which is
used to configure a multi-link scenario in PWEs, denoted as \noun{NNConfig}.
\begin{figure}[tp]
\centering{}\includegraphics[width=1\columnwidth]{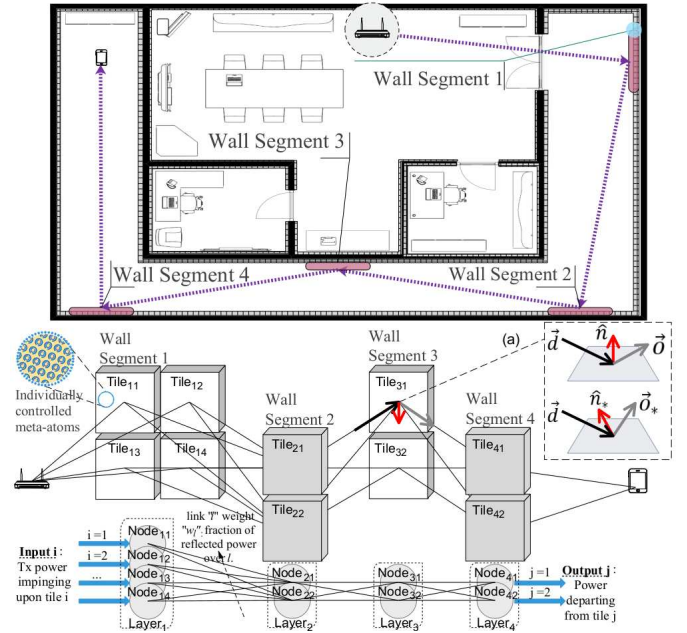}
\caption{\textcolor{black}{\label{fig:structure}(Top): A natural/free grouping
of complete SDM tile units is considered, e.g., per holding wall or
wall segment. Each tile has individually controlled meta-atoms. A
path of SDM groups to be traversed by TX-emitted waves is selected.
(Middle): (a) Equivalent, virtual rotation of tiles, to match an impinging
direction $\vec{d}$ to a required reflection direction $\vec{o}$.
(Bottom) Neural network representation of the PWE.}}
\end{figure}

We assume an environment with $N\left(\geq2\right)$ pairs of transmitters
(TXs) and receivers (RXs), with each link denoted with an index $n={1,2,...,N}$.
With no loss of generality, all direct links from TXs to RXs are obstructed,
which means that there is no line-of-sight (LOS) path among each pair
of TX and RX, in order to focus our study on the configuration of
the PWEs to maximize \textcolor{black}{their utility. In each wall,
the number of tiles, denoted as $\left\Vert \textrm{Wall}_{i}\right\Vert $
is generally larger than $N$ (i.e., $\left\Vert \textrm{Wall}_{i}\right\Vert >N$).
The subscript $i$ represents the index of the wall and $\left\Vert *\right\Vert $
is the cardinality of the set of tiles of the corresponding wall.
The set of tiles are expressed as $\{\textrm{Tile}_{ij}:i\in\mathbb{N},j\in\mathbb{N}\}$,
where $j$ is the index of tile of the $i$-th wall, with each tile
has a distinctly indexed.}

\subsection{Configuration of a PWE as a Fully Connected Neural Network}

\subsubsection{\textcolor{black}{Structuring a NN for PWEs}}

\textcolor{black}{The proposed neural network architecture sets as
a primary goal to be directly interpretable, as shown in Fig.~\ref{fig:structure}.
We define that: i) each neuron (hidden or not) will represent exactly
one physical SDM tile unit, and ii) the state of a neuron will directly
correspond to a parameterized tile EM functionality. }

\textcolor{black}{For point (i), we begin by defining the number of
layers and number of nodes in each layer. The number of layers corresponds
to the number of walls (which have SDM tile coating) that a TX-emitted
wave will sequentially impinge upon to reach an intended RX. Hence,
for each TX-RX pair, the first step is to select the two SDM-coated
walls in direct proximity of the TX and RX, respectively, which are
mapped to the input and output layers of the neural network.}

\textcolor{black}{For point (ii), we employ a NN architecture which
is qualitatively as close as possible to the actual physical propagation
phenomenon. Thus, since tiles manipulate the flow of EM waves from
a TX to an RX, we employ a feedforward NN since it enforces the same,
similar activation and actuation of its neurons. Backpropagation is
employed to re-tune the nodes (and corresponding tile functionalities)
based on the received power pattern at the output layer at the previous
feedforward step. A linear ramp will act as the node activation function,
in order to be aligned with energy conservation principle of wireless
propagation (i.e., the reflected power from a tile should be equal
to the impinging one, in the ideal case). Any link weight will represent
the portion of the tile-impinging energy that is reflected towards
the link direction.}

\textcolor{black}{We proceed to study and details these design principles
further.}

\subsubsection{Mapping a PWE into a NN}

\textcolor{black}{The mapping process of SDM tiles into neural network
nodes is visualized in Fig.~\ref{fig:structure}. First, we assume
a grouping of tiles into larger groups for scalability reasons. This
grouping can be defined freely. For instance, since SDM tiles cover
floorplan walls, one can naturally group together tiles that are placed
over the same wall. Second, an order (path) of SDM groups is selected
to serve as a coarse air-path connective the TX to the RX. This ordering
is derived via any path finding algorithm (e.g., shortest path), over
a graph comprising SDM groups as vertexes and SDM groups in LOS as
edges. An example is shown in Fig.~\ref{fig:structure}-top and Fig.~\ref{fig:structure}-middle
insets. Notice that the first group in this order is the one right
after the TX. This group is deterministically set as the group of
SDM tiles that receive TX-emitted waves impinging upon them. This
information can be derived via the direct sensing capabilities of
the SDM tile hardware~\cite{ABSENSE}. Thus, the number of tiles
in the first group and their identity is set. As a rule of a thumb,
all subsequent groups are assumed to contain an equal number of SDM
tile to the first layer. Finally, the ordering of groups is mapped
to the neural network shown in Fig.~\ref{fig:structure}-bottom inset.
Each group is mapped to a neural network layer, and each specific
SDM tile to a neural network node. A neural network link is inserted
for each SDM tile pair within LOS of each other. The first group is
the input layer and the last group is the output layer of the neural
network.}

\textcolor{black}{The selection of intermediate layer-walls follows
the principles of the $\textsc{KpConfig}$ approach~\cite{liaskos2019network},
summarized as follows. Since ideal SDMs can freely redirect any impinging
direction to any other, the wave propagation becomes akin to routing
in graph comprising SDM-coated walls as vertexes and links between
walls within LOS of each other. The selection of intermediate walls
follows a shortest-path finding procedure over this graph. All available
tiles in selected walls over this shortest path become nodes of the
corresponding wall-layer.}

\textcolor{black}{After determining the NN layers, the nodes/tiles
per wall are selected. It is assumed that when the electromagnetic
waves emitted from the TX impinge on the first wall, each tile in
the first wall receives certain power, which is a function of the
distance between the TX and the tile and the angle of arrival and
can be treated as the ``input'' of the neural network. All tiles
in the neural network can be tuned to redirect, split and focus its
impinging power to other walls and tiles in their line-of-sight, as
shown in black dashed lines in~Fig.~\ref{fig:structure}. The tiles
reflect waves in tunable elevation and azimuth planes (a detailed
discussion is provided below). After waves propagate via all wirelessly
``connected'' tiles, they reach the final wall-layer via multiple
paths. The reflections from the final wall to the UE (User Equipment)
is considered as the final ``output''. Ideally, the final output
received by the UE after reflections within the PWEs should be the
transmitted power minus any path losses and any SDM reflection losses
(which can be near-zero for ideal metasurfaces\ \cite{MSSurveyAllFunctionsAndTypes}).
Hence, we can obtain a metric of deviation which is also the cost
function of the proposed neural network, denoted as $\xi$, from the
ideal output and actual measured values through different configurations
of tiles. The root mean square error (RMSE) constitutes a common choice
for $\xi$.}

It is noted that metasurfaces can focus a wave as well as steer it
towards an intended direction, which a degree of efficiency that is
unique to metasurfaces over reflectarrays and antenna arrays. This
capability can counter-balance the free-space path loss with a corresponding
loss-canceling reflection gain. Thus, the overall path loss is essentially
defined by any power consumed over the metasurface materials per each
bounce~\cite{liaskos2019network}.
\begin{rem}
\label{rem:Essentially,-the-steering}Essentially, the steering of
a wave impinging on an SDM tile from a direction $\vec{d}$ to a reflected
direction $\vec{r}$, is modeled as inducing a corresponding alteration
to the vector $\hat{n}$, as shown in Fig.~\ref{fig:structure}-(a).
This convention was introduced in \cite{liaskos2019network} and is
denoted as \emph{virtual rotation} of the normal $\hat{n}$. This
means that the physical orientation of the SDM is not altered, but
rather the SDM configuration acts as if the tile was rotated in a
corresponding fashion to cause the intended reflection.
\end{rem}
Note that the signals impinging on a tile can be scattered to more
than one directions, depending on directions of arrival and the configuration
of the tile. We describe this partial scattering of power over a set
of reflection directions $\overrightarrow{o}$ using a heuristic power
fraction $w_{\vec{o}}\left(\hat{n},\vec{d}\right)$, expressed as:
\begin{equation}
w_{\vec{o}}\left(\hat{n},\vec{d}\right)=\frac{\max\left\{ \vec{r}\left(\hat{n},\vec{d}\right)\cdot\vec{o},0\right\} }{\sum_{\forall\vec{o}}w_{\vec{o}}},\label{eq:weights-1}
\end{equation}
in which $\vec{o}$ is the outgoing direction, $\vec{r}$ the reflecting
direction, $\vec{d}$ the impinging direction. The projection of $\vec{r}$
over $\vec{o}$ is normalized across all outgoing links, assuming
that $\vec{o}$ and $\vec{d}$ are unit vectors. A negative inner
product implies that vectors $\vec{r}$, $\vec{o}$ are not aligned
in the same direction and, thus, $w_{\vec{o}}$ should intuitively
be zero.

According to the comparison result, the virtual rotation of the normal
of each tile defines the reflected wave direction, obtained via the
standard rule for calculating a reflection $\vec{r}$ from an impinging
vector $\vec{d}$ and a surface normal $\hat{n}$:
\begin{equation}
\vec{r}\left(\hat{n},\vec{d}\right)=\vec{d}-2\left(\vec{d}\cdot\hat{n}\right)\hat{n},\label{eq:lawReflect}
\end{equation}
Thus, $\hat{n}$ can be updated to minimize the c\textcolor{black}{ost
function $\xi$:
\begin{equation}
\xi\left(\hat{n},\vec{d}\right)=\frac{1}{2}\sum_{l=1}^{L}\left(\bar{p}-\rho_{l}\left(\hat{n},\vec{d}\right)\right)^{2},
\end{equation}
where $\bar{p}$ is the ideal output power value which is equal for
each tile, $\rho_{l}$ the actual output power of the $l$-th tile
within a wall of $L$ tiles, and $\rho_{l}\left(\hat{n},\vec{d}\right)=w_{\vec{o}}\left(\hat{n},\vec{d}\right)\cdot\bar{p}$.}

\subsection{Rotation Matrices in PWEs\label{sec:rotation}}

In order to describe a rotation of an object from its original orientation,
several approaches are known: Euler angles, rotation axis and angle,
quaternions, and rotation matrix~\cite{dorst2010geometric}. Among
all these methods, the rotation matrix is the most suitable for the
tile rotations in PWEs, since even though it has the expression of
$3\times3$ matrices, it only introduces one parameter per plane,
i.e., a rotation angle, to describe the rotation in the space. Additionally,
since the tile normals are unary vectors, the effective degree of
freedom for their virtual rotation is reduced to two (i.e., an azimuth
and elevation defining the end-point of $\hat{n}$ over a sphere with
unary radius, and an origin at the tile center). Therefore, to quantify
the rotation of tiles in Euclidean space, we introduce the rotation
matrices from linear algebra in a three-dimensional Cartesian coordinate
system. In cases with no rotation, the normal vector of a plane $\hat{n}$
after normalization are:
\begin{equation}
\begin{aligned}\hat{n}_{\textsc{xz}}^{\pm\textsc{y}} & =[0\quad\pm1\quad0]^{\intercal},\\
\nonumber\hat{n}_{\textsc{yz}}^{\pm\textsc{x}} & =[\pm1\quad0\quad0]^{\intercal},\\
\nonumber\hat{n}_{\textsc{xy}}^{-\textsc{z}} & =[0\quad0\quad-1]^{\intercal},
\end{aligned}
\end{equation}
where the subscript of each $\hat{n}$ indicates the plane which the
normal vector belongs to, and the superscript shows the pointing direction
of the normal vector. Based on observations from previous studies~\cite{liaskos2018new,liaskos2018realizing},
we make the following remark regarding the effective rotation matrices:
\begin{rem}
The tiles rotate virtually in only two planes, thus having two degrees
of freedom. A tile fixed at the $xz$ plane has effective rotations
around the $x$ and $z$ axes only.
\end{rem}
The rotation matrices $R_{XZ},\,R_{YZ},R_{XY}$ are\textcolor{black}{:}\textcolor{black}{\scriptsize{}
\begin{multline}
\begin{bmatrix}\cos\phi & -\sin\phi & 0\\
\cos\theta\sin\phi & \cos\theta\cos\phi & -\sin\theta\\
\sin\theta\sin\phi & \sin\theta\cos\phi & \cos\theta
\end{bmatrix},\\
\begin{bmatrix}\cos\varphi\cos\phi & -\cos\varphi\sin\phi & \sin\varphi\\
\sin\phi & \cos\phi & 0\\
-\sin\varphi\cos\phi & \sin\varphi\sin\phi & \cos\varphi
\end{bmatrix},\\
\begin{bmatrix}\cos\varphi & 0 & \sin\varphi\\
\sin\theta\sin\varphi & \cos\theta & -\sin\theta\cos\varphi\\
-\cos\theta\sin\varphi & \sin\theta & \cos\theta\cos\varphi
\end{bmatrix}
\end{multline}
}\textcolor{black}{respectively, where $\theta$, $\phi$, and $\varphi$
are the angles of rotation around the x, z, and y axis. It is }worth
noting that the rotation matrix is actually readily extendable to
planar walls which might not be perpendicular to any Cartesian plane.
The straightforward solution is to multiply the walls' own orientation
matrix to the tiles, then the rest of the procedure holds. Having
a quantifiable metric for tile rotations $\hat{n}$, the power fraction
calculation can be rewritten as
\begin{equation}
w_{\vec{o}}\left(\theta,\phi,\varphi,\vec{d}\right)=\frac{\max\left\{ \vec{r}\left(\theta,\phi,\varphi,\vec{d}\right)\cdot\vec{o},0\right\} }{\sum_{\forall\vec{o}}w_{\vec{o}}}.\label{eq:weights}
\end{equation}

\begin{rem}
In accordance with Remark \ref{rem:Essentially,-the-steering}, the
reflection $\vec{r}\left(\theta,\phi,\varphi,\vec{d}\right)$ in relation
(\ref{eq:weights}) is defined by a virtual rotation of the unit surface
normal $\hat{n}$. In a 3D setting, any such virtual rotation can
be obtained just from an azimuth and elevation angle relative to the
actual normal. As such, any two of the $\theta,\phi,\varphi$ angles
can be selected for the operation of the described neural network,
even on a per tile/node basis.
\begin{figure}[tp]
\centering{}\includegraphics[width=0.9\columnwidth]{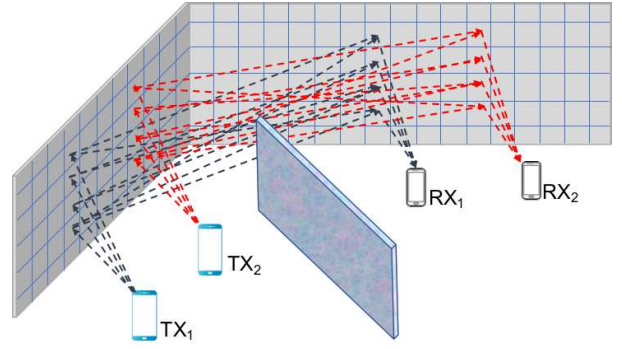}
\caption{\label{fig:multiLink}Multi-link PWEs with two pairs of TX and RX.
No line-of-sight link can be found between the pairs.}
\end{figure}
\end{rem}

\subsection{Ensemble Neural Networks for Multi-link PWEs\label{sec:ensembleNN}}

In PWEs with a multi-link scenario comprised of multiple walls, multiple
pairs of TX and RX can be served with minimal interference. Suppose
in a multi-link scenario, as illustrated in~Fig.~\ref{fig:multiLink},
the $l$-th tile which is located at the $k$-th wall and serves the
$n$-th link has the power of the reflecting EM waves as:
\begin{equation}
\rho_{j}^{(k,l,n)}=\sum_{\forall\vec{d}_{i}}w_{\vec{o}}\left(\theta_{i,n},\phi_{i,n},\varphi_{i,n},\vec{d}_{i}\right)\cdot P_{i}^{(k,l,n)},
\end{equation}
which is the feeding input power to the next layer (i.e., wall). At
the final layer ($k=K$), the metric of deviation $\xi_{n}$, is calculated
as:
\begin{equation}
\xi_{n}=\frac{1}{2}\sum_{\forall l}\left(\bar{\rho}^{(k,l,n)}-\rho^{(k,l,n)}\right)^{2}=\frac{1}{2}\sum_{\forall l}\delta_{l}^{2},
\end{equation}
where $\bar{\rho}$ denotes the ideal output power ratio, and:
\begin{equation}
\delta_{l}=\bar{\rho}^{(k,l,n)}-\rho^{(k,l,n)}.
\end{equation}
The reason for using ensemble neural networks for multi-link case
is that ensemble learning can reduce the variance of predictions and
minimize the error from multiple neural network models. The ensemble
learning approach can be grouped by elements with variable values,
such as different rotation matrices, number of tiles (nodes), and
various activation functions. The nonlinear activation function can
be hyperbolic tangent, softmax, or rectified linear unit (ReLU). Inside
the ensemble structure, there are multiple parallel neural networks
which follow the backpropagation rules to update the weights. After
the feedforward links between wall-layers have been iterated until
the final layer, each tile-node at each wall-layer (in reverse order)
deduces the effect of its current rotation angles $\theta,\phi,\varphi$
to the deviation $\xi$, and updates $\theta,\phi,\varphi$ to a new
set of values $\theta_{\ast},\phi_{\ast},\varphi_{\ast}$ . The corresponding
update rule can be generalized as{\small{}
\begin{equation}
\begin{aligned}\theta_{\ast}^{(k,l,n)} & =\theta^{(k,l,n)}-\eta\cdot\left(\frac{\partial\xi}{\partial\theta}\right)^{(k,l,n)}\cdot\mathcal{S}^{(k,l,n)},\\
\nonumber\phi_{\ast}^{(k,l,n)} & =\phi^{(k,l,n)}-\eta\cdot\left(\frac{\partial\xi}{\partial\phi}\right)^{(k,l,n)}\cdot\mathcal{S}^{(k,l,n)},\\
\nonumber\varphi_{\ast}^{(k,l,n)} & =\varphi^{(k,l,n)}-\eta\cdot\left(\frac{\partial\xi}{\partial\varphi}\right)^{(k,l,n)}\cdot\mathcal{S}^{(k,l,n)},
\end{aligned}
\end{equation}
w}here $\eta\in(0,1]$ is the network's learning rate, and $\mathcal{S}^{(k,l,n)}$
is a factor denoting the significance of tile $(k,l)$ in the $n$-th
link.

Note that given the remark made earlier, each update does not change
all three rotation angles: only two of them will be updated based
on the tile's location. For the final wall-layer $k=K$ we define
that $\mathcal{S}^{(K,l,n)}=\delta_{l}$, representing its deviation
from the local ideal output. For $k\neq K$ we define it as the total
power impinging on the tile, which is $\mathcal{S}^{(k,l,n)}=\sum_{\forall i}P_{i}^{(k,l,n)}$.

\emph{Inputs/Outputs}. The input to each tile of the first layer is
the normalized portion of power impinging upon it via the $\overrightarrow{d}$
links of user TX. Since the objective is to transfer all emitted power
to user RX, these inputs can be \emph{virtual} (i.e., not equal to
the actual impinging power distribution). Thus, each input is set
to $\nicefrac{1}{K}$ of the total emitted power, for each of the
$K$ tiles in the first layer (i.e., the first ``wall'' after the
TX). Using the same principle, the ideal output is set to $\nicefrac{1}{K^{*}}$,
with $K^{*}$ being the number of tiles in the last layer (i.e., before
the RX). Since the inputs and outputs remain the same at every feed-forward/back-propagate
cycle, a high learning rate can be selected (e.g., $\eta=0.95$).

\emph{Implementing the feedforward / backpropagate process}. The work
of~\cite{SPAWC.2019} introduced the formulation of $\left(\frac{\partial\xi}{\partial\phi}\right)^{(k,l,n)}$
in the $\xi$: RMSE case. The formulation yielded a custom feedforward
/ backpropagate implementation that operates at each node/tile, updating
the corresponding virtual normal $\hat{n}\left(\phi,\theta\right)$
based on the status of the right-hand connected nodes. The corresponding
right-hand neural links are then set based on relation (\ref{eq:weights}).

However, it is noted that this custom implementation approach is not
be directly compatible with the existing array of neural network software
and hardware suites, which operate at the link-level, rather than
at the node-level. Therefore, in this iteration of the scheme we propose
an adaptation that provides compatibility with feedforward / backpropagate
implementations operating at the link-level. Assuming any link update
process, and once the corresponding right-hand link weights\textendash $\mathring{w_{\overrightarrow{o}}}$\textendash of
a node have been updated, we introduce the following steps:
\begin{itemize}
\item From relation (\ref{eq:weights}), calculate the virtual normal rotation,
$\hat{n^{*}}$, that best matches the updated right-hand weights,
i.e.,
\begin{equation}
\hat{n^{*}}\gets\text{argmin}_{\hat{n}}\xi\left(w_{\vec{o}}\left(\hat{n},\vec{d}\right)-\mathring{w_{\overrightarrow{o}}}\right),
\end{equation}
where $\xi$ is an error norm such as the RMSE, while $w_{\vec{o}}\left(\hat{n},\vec{d}\right)$
is given in relation~(\ref{eq:weights-1}). Notice that the match
is approximate in the general case.
\item Then, we proceed by updating anew the right-hand weights of the node
via relation (\ref{eq:weights}), using the calculated value of $\hat{n^{*}}$.
\end{itemize}
Notice that weight adaptations are not uncommon in neural networks
in general. For instance, weight freezing\textendash the process of
excluding certain links from the weight update process\textendash can
be implemented by calculating a new link weight in the bulk of the
training cycle, but eventually not updating its previous value~\cite{islam2001new}.

These feed-forward/back-propagate cycles in each neural network can
be executed in an online or offline manner, the ensemble learning
process will finish until the deviation $\xi$ arrives at its global
optimum, reaches an acceptable level or an allocated computational
time window expires. At this point, the PWE controller simply deploys
EM functionalities at each tile $(k,l)$ of each link, matching the
attained $w_{\vec{o}}\left(\theta,\phi,\varphi,\vec{d}\right)$ values.\textcolor{blue}{}
\begin{figure}[tp]
\centering{}\includegraphics[width=0.9\columnwidth]{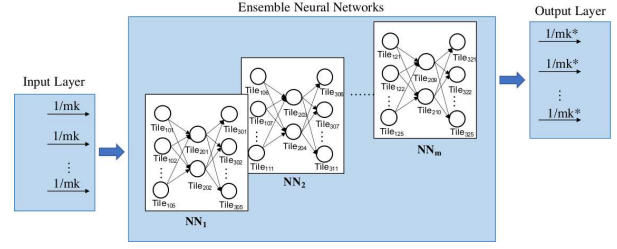}
\caption{\label{fig:ENN}Structure of the proposed ensemble neural networks
to enable global optimization over a large number of tiles/nodes.}
\end{figure}

\textcolor{black}{\emph{Constructing Ensemble Neural Networks.}}\textcolor{black}{{}
In cases where a single layer-wall is too large to construct a single
neural network with all nodes fully connected, an ensemble neural
network can relieve the computational burden. As shown in~Fig.~\ref{fig:ENN},
a total of $m$ fully connected neural networks are constructed with
subsets of available tiles from wall-layer. The same input values
(i.e., $1/(mk)$ due to the presence of $m$ neural networks) are
injected to each node in the first hidden layer of each neural network.
In the each neural network, a subset of tiles are selected for training.
For example, in~Fig.~\ref{fig:ENN}, each of the first and third
layer has 25~tiles while the second has 10, a total of $m=5$ neural
networks constitute an ensemble neural network to search for the optimal
configuration of each tile~}\textcolor{black}{\noun{NNConfig}}\textcolor{black}{.
Accordingly, the desired output value from each output layer becomes
$1/(mk^{\ast})$. }

\subsection{Interpretability: Mapping a trained neural network to tile configurations\label{subsec:Interpretability:-Mapping-a}}

As described in the context of relation~(\ref{eq:weights}), the
feedforward process of \noun{NNConfig} is not an accurate representation
of the wave propagation process. Instead, for the sake of having a
simplified computational rule that runs on each node of the neural
network, relation~(\ref{eq:weights}) enforces a link weight derivation
based on a simple\emph{ projection} of a wave reflection (derived
from an incoming direction, $\vec{d}$, and the virtual tile surface
normal, $\hat{n}$), over each outgoing direction, $\vec{o}$. Therefore,
once the \noun{NNConfig} network has been trained, its outcomes must
be mapped to actual tile configurations that reflect the expected
propagation outcome.

To this end, we consider a single trained neural network node and
its corresponding tile. Let:
\begin{equation}
\vec{d}_{i},\,i=1\ldots D:\,w\left(\vec{d}_{i}\right)\ne0
\end{equation}
be the set of tile links that carry wireless waves with non-zero power
that impinge over the tile, according to the \noun{NNConfig} training
outcome. Similarly, let:
\begin{equation}
\vec{o}_{i},\,i=1\ldots O:\,w\left(\vec{o}_{i}\right)\ne0
\end{equation}
be the set of tile links that carry wireless waves with non-zero power
that depart from the tile. Additionally, let $\left\Vert *\right\Vert $
be the cardinality of a set $*$. Then, the following cases are defined:
\begin{itemize}
\item \textbf{Case} $\left(\left\Vert \vec{d}_{i}\right\Vert =1\right)\,\&\,\left(\vphantom{\vec{d}_{i}}\left\Vert \vec{o}_{i}\right\Vert =1\right)$.
This case reflects a simple wave steering scenario, where a wave impinging
from direction $\vec{d}_{i}$ is redirected to direction $\vec{o}_{i}$,
with a corresponding virtual normal $\hat{n}$. Thus, the interpreted
tile function can be written as:
\begin{equation}
\text{\textsc{Steer}}\left(\vec{d}_{1},\vec{o}_{1}\right)\to\hat{n}.\label{eq:steer}
\end{equation}
\item \textbf{Case} $\left(\left\Vert \vec{d}_{i}\right\Vert \ge1\right)\,\&\,\left(\vphantom{\vec{d}_{i}}\left\Vert \vec{o}_{i}\right\Vert =0\right)$.
In this case, waves impinge upon the tile from one or more directions
while no reflections are requested, yielding a wave absorption function:
\begin{equation}
\text{\textsc{Absorb}}\left(\vec{d}_{i}\right).
\end{equation}
From a physical standpoint of view, it is noted that wave absorption
is perfect (i.e., yielding practically near-zero reflections) only
when $\left\Vert \vec{d}_{i}\right\Vert =1$~\cite{MSSurveyAllFunctionsAndTypes}.
When $\left\Vert \vec{d}_{i}\right\Vert >1$, absorption may be partially
effective and lead to a degree of wave scattering. In that case, we
apply:
\begin{equation}
\text{\textsc{Absorb}}\left(\vec{d}_{I}\right),\,I:argmax_{i}\left(w\left(\vec{d}_{i}\right)\right),
\end{equation}
i.e., configuring the tile for absorbing perfectly the strongest impinging
wave.
\item \textbf{Case} $\left(\left\Vert \vec{d}_{i}\right\Vert =1\right)\,\&\,\left(\vphantom{\vec{d}_{i}}\left\Vert \vec{o}_{i}\right\Vert >1\right)$.
In this case, a single impinging wave is reflected towards multiple
directions, each carrying a portion of the original power. This functionality
is known as splitting, i.e.:
\begin{equation}
\text{\textsc{Split}}\left(\vec{d}_{1},\left\langle \vec{o}_{i},w\left(\vec{o}_{i}\right)\right\rangle \right)\label{eq:split}
\end{equation}
\item \textbf{Case} $\left(\left\Vert \vec{d}_{i}\right\Vert >1\right)\,\&\,\left(\vphantom{\vec{d}_{i}}\left\Vert \vec{o}_{i}\right\Vert >1\right)$.
This constitutes the most general wave scattering case, which may
not always have a unique mapping to a tile functionality. For instance,
this scattering may be attributed to a splitting function (such as
the one in relation~(\ref{eq:split})), when the tile is also illuminated
by waves incoming from directions other than the intended one. Similarly,
another possible cause of this scattering is to have a single steering\textendash as
defined in equation~(\ref{eq:steer})\textendash while the tile gets
illuminated once again by one or more unintended incoming directions.
Since the main goal of the proposed \noun{NNConfig }is to use a single
tile functionality for multiple connectivity objectives, this case
will be mapped based on two principles: i) ensure at least one strong
connection from set $\vec{d}_{i}$ to set $\vec{o}_{i}$, ii) pick
the one specific connection that best fits the total sets $\vec{d}_{i}$
to $\vec{o}_{i}$. This is expressed as:
\begin{equation}
\exists\hat{n}:\,\forall\vec{d}\in O_{d}\left(\vec{d_{i}}\right),\forall\vec{o}\in O_{o}\left(\vec{o_{i}}\right):\text{\textsc{Steer}}\left(\vec{d},\vec{o}\right)\to\hat{n}.\label{eq:multiSteer}
\end{equation}
In other words, there exists a steering functionality that reflects
each indexed element of $O_{d}\left(\vec{d_{i}}\right)$ to the element
of $O_{o}\left(\vec{o_{i}}\right)$ with the same index, for some
orderings $O_{d}$, $O_{o}$ of the original sets.
\begin{algorithm}[!t]
\begin{algorithmic}[1]
\Procedure{\noun{\small{}MultiSteerMap}}{{${\vec{d}}_{i}$,${\vec{o}}_{i}$}}
\State {\small{}$best\_score\gets\infty;$}{\small\par}
\State {\small{}$best\_pair\gets\emptyset;$}{\small\par}
\State \textbf{\small{}for }{\small{}$\overrightarrow{d}$}\textbf{\small{} in}{\small{} $\overrightarrow{d}_{i}$}{\small\par}
\State ~~\textbf{\small{}for}{\small{} $\overrightarrow{o}$}\textcolor{black}{\small{} }\textbf{\textcolor{black}{\small{}in}}\textcolor{black}{\small{} $\overrightarrow{o}_{i}$}{\small\par}
\State ~~\textbf{~~}{\small{}$\hat{n}{\scriptscriptstyle \gets}\text{\textsc{Steer}}\left(\vec{d},\overrightarrow{o}\right)$;}{\small\par}
\State ~~~~{\small{}$Set\,S\gets\text{\textsc{CopyOf}}\left(\overrightarrow{o}_{i}\right);$}{\small\par}
\State ~~\textbf{\small{}~~for }{\small{}$\overrightarrow{\delta}$}\textbf{\small{} in}{\small{} $\overrightarrow{d}_{i}$}{\small\par}
\State ~~\textbf{~~~~}\textbf{\small{}$S\gets S-\left(\overrightarrow{\delta}-2\left(\overrightarrow{\delta}\cdot\hat{n}\right)\hat{n}\right);$}{\small\par}
\State ~~\textbf{~~}\textbf{\small{}end for}{\small\par}
\State ~~~~\textbf{\small{}if}{\small{} $\left\Vert S\right\Vert <best\_score$ }\textbf{\small{}then}{\small\par}
\State ~~~~~~\textbf{\small{}$best\_score\gets\left\Vert S\right\Vert ;$}{\small\par}
\State ~~~~~~\textbf{\small{}$best\_pair\gets\left\langle \overrightarrow{d},\overrightarrow{o}\right\rangle $}{\small\par}
\State ~~\textbf{\small{}~~end if}{\small\par}
\State ~~\textbf{\small{}end for}{\small\par}
\State \textbf{\small{}end for}{\small\par}
\State return $best\_pair${\small{}; }{\small\par}
\EndProcedure
\end{algorithmic}

\caption{\label{alg:multiSteer}The multi-steer interpretation process.}
\end{algorithm}
\end{itemize}
The definition of mapping~(\ref{eq:multiSteer}) is strict, in the
sense that there may not exist an $\hat{n}$ that meets it precisely.
Therefore, the \noun{MultiSteerMap} process (Algotihm~\ref{alg:multiSteer})
presents a practical, approximate heuristic. The algorithm goes through
all possible incoming/outgoing direction pairs, and considers the
prospects of deploying the corresponding \noun{Steer} function. The
pair that leads to the connectivity of the most $\vec{d}_{i}$ to
$\vec{o}_{i}$ elements is finally selected.

For the sake of completion we also mention the case where $\left(\left\Vert \vec{d}_{i}\right\Vert =0\right)\,\&\,\left(\vphantom{\vec{d}_{i}}\left\Vert \vec{o}_{i}\right\Vert \ge1\right)$,
i.e., a tile with zero impinging power and non-zero reflected one,
which is an invalid case. Note that this trained node outcome is naturally
prohibited from the training process, due to the form of relation~(\ref{eq:weights}).

As explained in~\cite{liaskos2019network}, certain metasurface functionalities
are disjoint and independent from others. The most critical functionality
is to steer RF power and guide it to an RX. This is accomplished by
the proposed scheme, citing the more economic use of tiles versus
the scheme of~\cite{liaskos2019network}, i.e., using fewer tiles
to serve a given set of users. The output is one or more ``air-routes''
connecting the emissions of a TX to an RX.

Subsequent metasurface functionalities, i.e., polarization alteration
and phase alteration can be applied to any ``air-route'' at the
final step of the PWE configuration. Just one tile per ``air-path''
can be used to introduce a phase shift or a polarization rotation
per air-path, without affecting the steering outcome deduced by the
proposed NNConfig, as discussed in~\cite{liaskos2019network}. We
note that, in this manner, phase and polarization control can be set
deterministically (e.g., to cancel fading effects) and they need not
be subject to a heuristic optimizer such as a neural network.

Regarding the frequency control, metasurfaces can indeed act as frequency
filters, but are usually not tunable in that sense. In other words,
when activated, a metasurface filters impinging frequencies, while
it does not do so when remaining inactive. As such, this functionality
is not taking into account into the neural network either.

Finally, regarding the obstacle avoidance, NNConfig receives as input
a series of walls that form a path which bypasses LOS obstacles. (Note
that avoiding malevolent users and protecting floorplan areas from
interference can be treated similarly). This series of walls is derived
via path-finding algorithms (e.g., K-Shortest Paths~\cite{liaskos2019network})
within a graph constructed as follows: walls are mapped to graph nodes
and walls within LOS of each other are mapped to graph-links. This
path-finding functionality is already used in KpConfig~\cite{liaskos2019network}
and is re-used in this work as input.

Ultimately, the approximate TX/RX user device position is treated
as an input provided by external services (as shown in Fig.~\ref{fig:SDM}
and its discussion). The neural network training criterion is\textcolor{black}{{}
the }RX received power, and as such any uncertainty in the device
locations is handled automatically by the neural network training
process to the extend possible.

\section{Evaluation\label{sec:Evaluation}}

We evaluate the performance of \noun{NNConfig} in the PWE simulator
presented in~\cite{liaskos2019network}. We seek to evaluate the
potential of the proposed scheme in a ray-tracing setting, from the
aspects of: i) receiver signal power levels, ii) neural network training,
and iii) tile numbers usage.

We compare its outcomes to the KpConfig scheme for PWEs presented
in~\cite{liaskos2019network} and whose operation has been outlined
in Section~\ref{sec:Related-work}. To the best of our knowledge,
KpConfig is the only scheme that operates at the PWE configuration
layer (i.e., definition of wave steering commands) and, therefore,
directly comparable to the proposed NNConfig.

\begin{figure}[tp]
\begin{centering}
\subfloat[\label{fig:KpPropag2wall}\noun{KpConfig} propagation.]{\begin{centering}
\includegraphics[width=0.45\columnwidth]{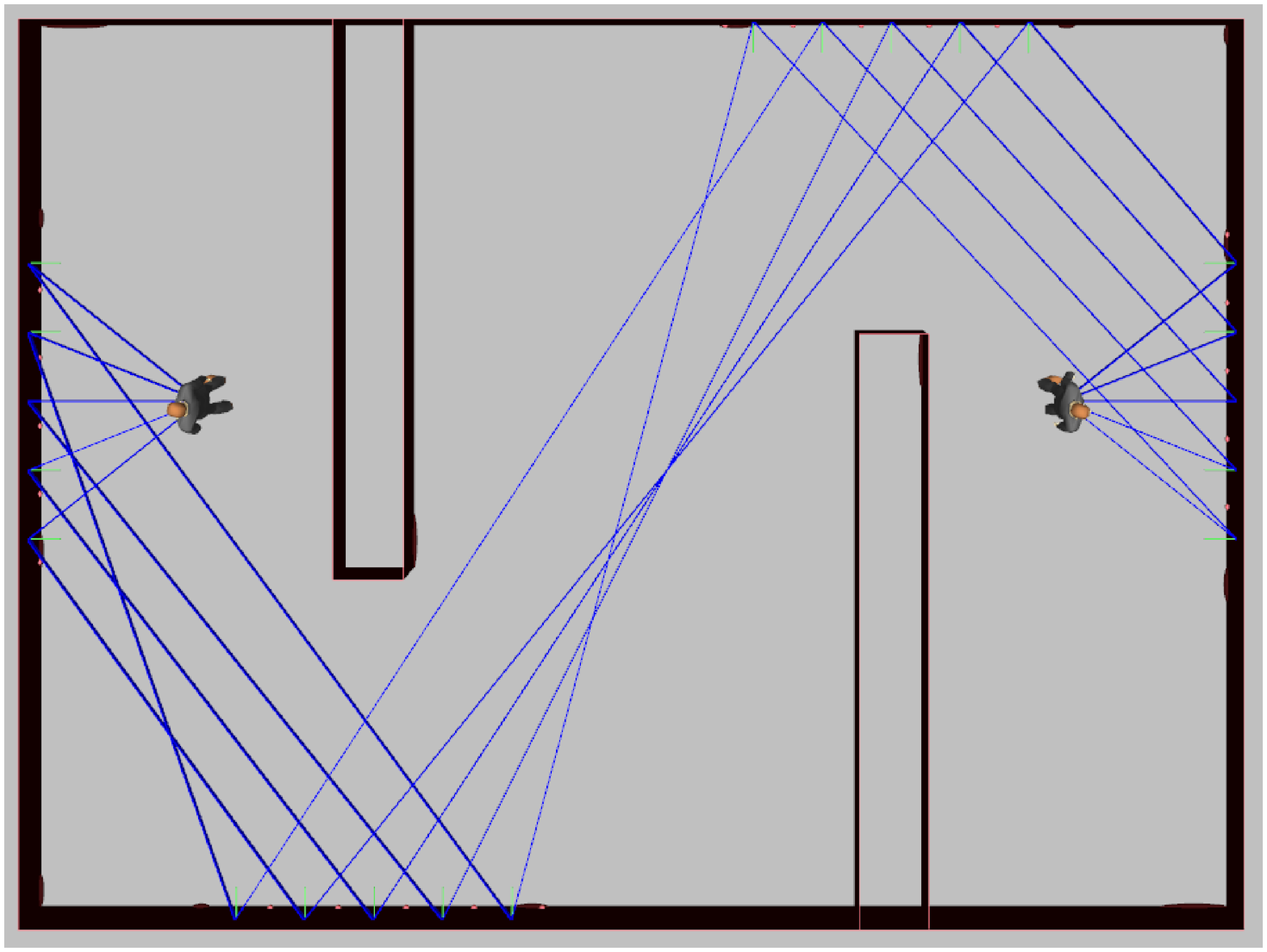}
\par\end{centering}
}\subfloat[\label{fig:NnPropag2wall}\noun{NNConfig} propagation.]{\begin{centering}
\includegraphics[width=0.45\columnwidth]{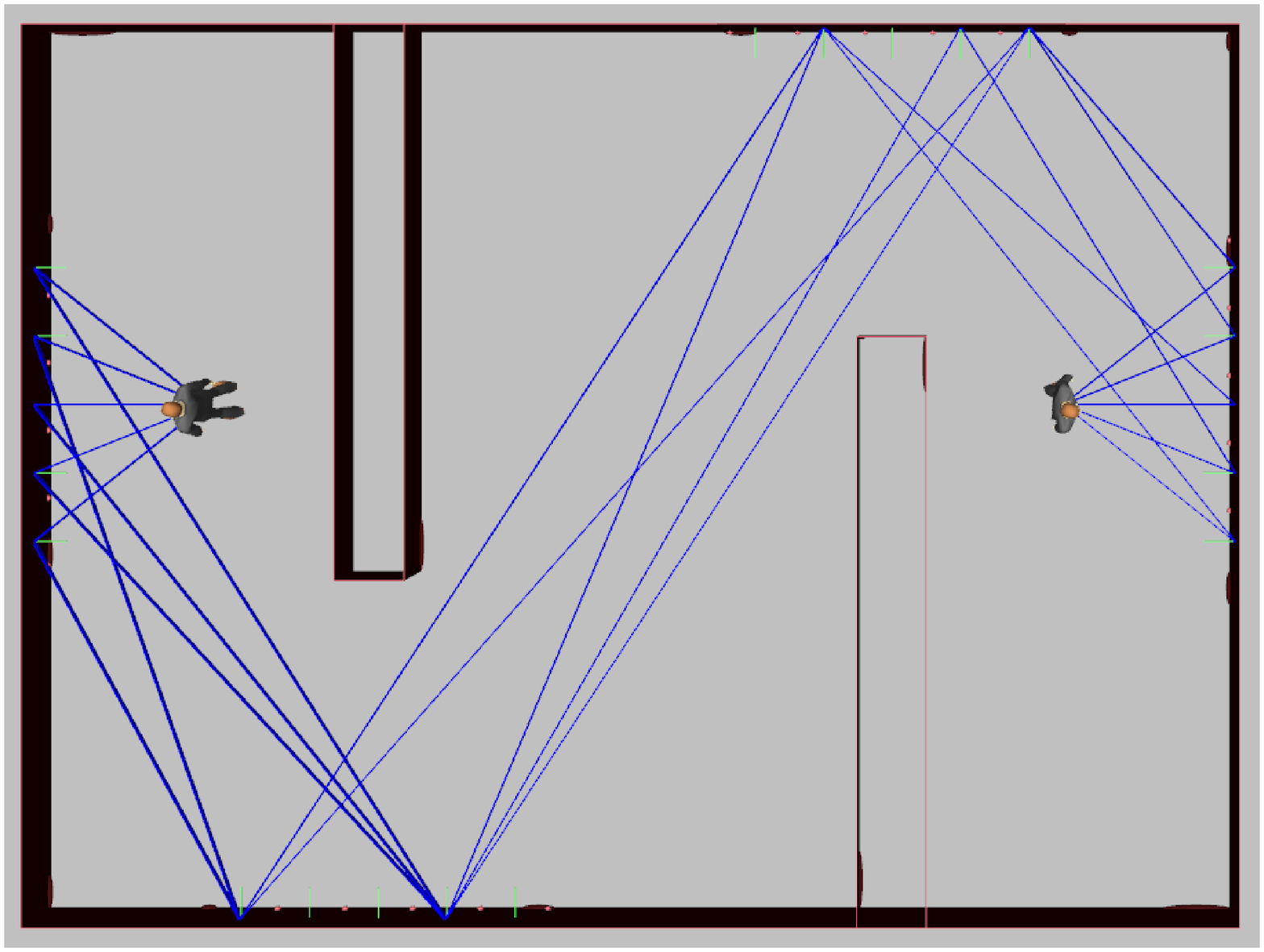}
\par\end{centering}
}
\par\end{centering}
\caption{\label{fig:Propag2walls}\noun{KpConfig }and\noun{ NNConfig }behavior
at the 2-middle wall\textcolor{black}{{} scenario. The utilization of
}\textcolor{black}{\noun{NNConfig}}\textcolor{black}{~maintains full
link connectivity while reducing deployment of five tiles, compared
with the same link using~}\textcolor{black}{\noun{KpConfig}}\textcolor{black}{.}}
\end{figure}

As explained in the context of Fig.~\ref{fig:structure}, NNConfig
receives as input a series of tile sets (walls), which act as the
neural network layers. (In the context of this evaluation, this is
accomplished by executing KpConfig, which produces end-to-end tile-disjoint
paths between the RX and the TX, as shown in Fig.~\ref{fig:KpPropag2wall}.
Any KpConfig-derived tile functionality is then discarded, and we
ke\textcolor{black}{ep only the same-wall tiles, which act as the
NNConfig nodes and layers, as shown in Fig.~\ref{fig:structure}).
In order to evaluate the NNConfig for varying number of hidden layer
cases, we consider the floorplan scenarios shown in Table~\ref{tab:Floorplan}.
In scenario~1, propagating waves necessarily need to impinge upon
1 hidden layer (i.e., NLOS wall), with the LOS walls of the TX and
the RX acting as the input and output layers respectively, producing
a neural network with a total of 3 layers, i.e., $5\times5\times5$
assuming 5 tiles per wall. Similarly, we have a NN of $5\times5\times5\times5$
for scenario 2, $5\times5\times5\times5\times5$ for scenario 3, $5\times5\times5\times5\times5\times5$
for scenario 4, and $5\times5\times5\times5\times5\times5\times5$
for scenario 5. In general, scenario~$I$ corresponds to $I$ hidden
neural network layers and $I+2$ layers total. The tile numbers within
each layer are defined as $\left(\text{tiles\,over\,wall-layer}\right)\times\left(\text{NN pruning factor}\right)$,
where NN pruning is a factor introduced for experimenting with a variable
ensemble neural network structure. Apart from this variable change
in the node numbers per layer, all other ensemble parameters are identical
to those defined above.}

\begin{table}[t]
\centering{}\caption{\label{tab:Floorplan}\noun{Floorplan, user location and beam orientation
coordinate system. The origin is at the lower left corner of the floorplan}.}
\begin{tabular}[b]{cc}
\begin{tabular*}{3.8cm}{@{\extracolsep{\fill}}@{\extracolsep{\fill}}@{\extracolsep{\fill}}@{}l}
\multicolumn{1}{l}{\textsc{User: Position, $\alpha$,$\phi$,$\theta$}}\tabularnewline
\midrule
TX(0): {[}2.5, 7.5, 1.5{]}, 40$^{o}$, 0$^{o}$, 0$^{o}$ \tabularnewline
RX(1): {[}$L$-2.5, 7.5, 1.5{]}, 40$^{o}$, 0$^{o}$, 180$^{o}$ \tabularnewline
~~~~$L$:~\noun{floorplan total width}\tabularnewline
\midrule
\multicolumn{1}{l}{\textsc{Comm. Pair: }\textcolor{black}{TX(0)$\to$RX(1)} }\tabularnewline
\midrule
\textsc{Objective}: $\text{\textsc{Max. RX Power}}$\tabularnewline
\bottomrule
\end{tabular*} & %
\begin{tabular}{c}
\tabularnewline
\hspace{35bp}\includegraphics[width=0.2\columnwidth]{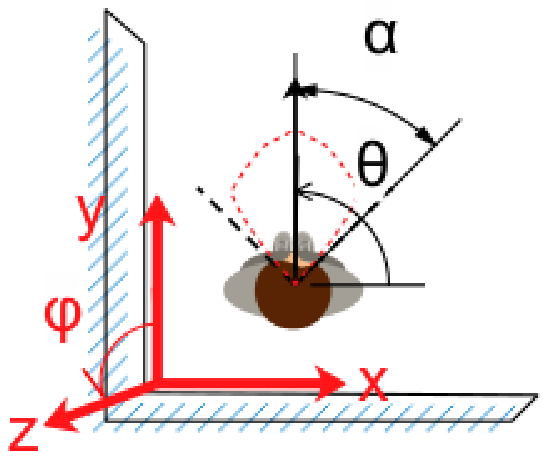}\tabularnewline
\end{tabular}\tabularnewline
\multicolumn{2}{c}{\includegraphics[width=0.14\columnwidth]{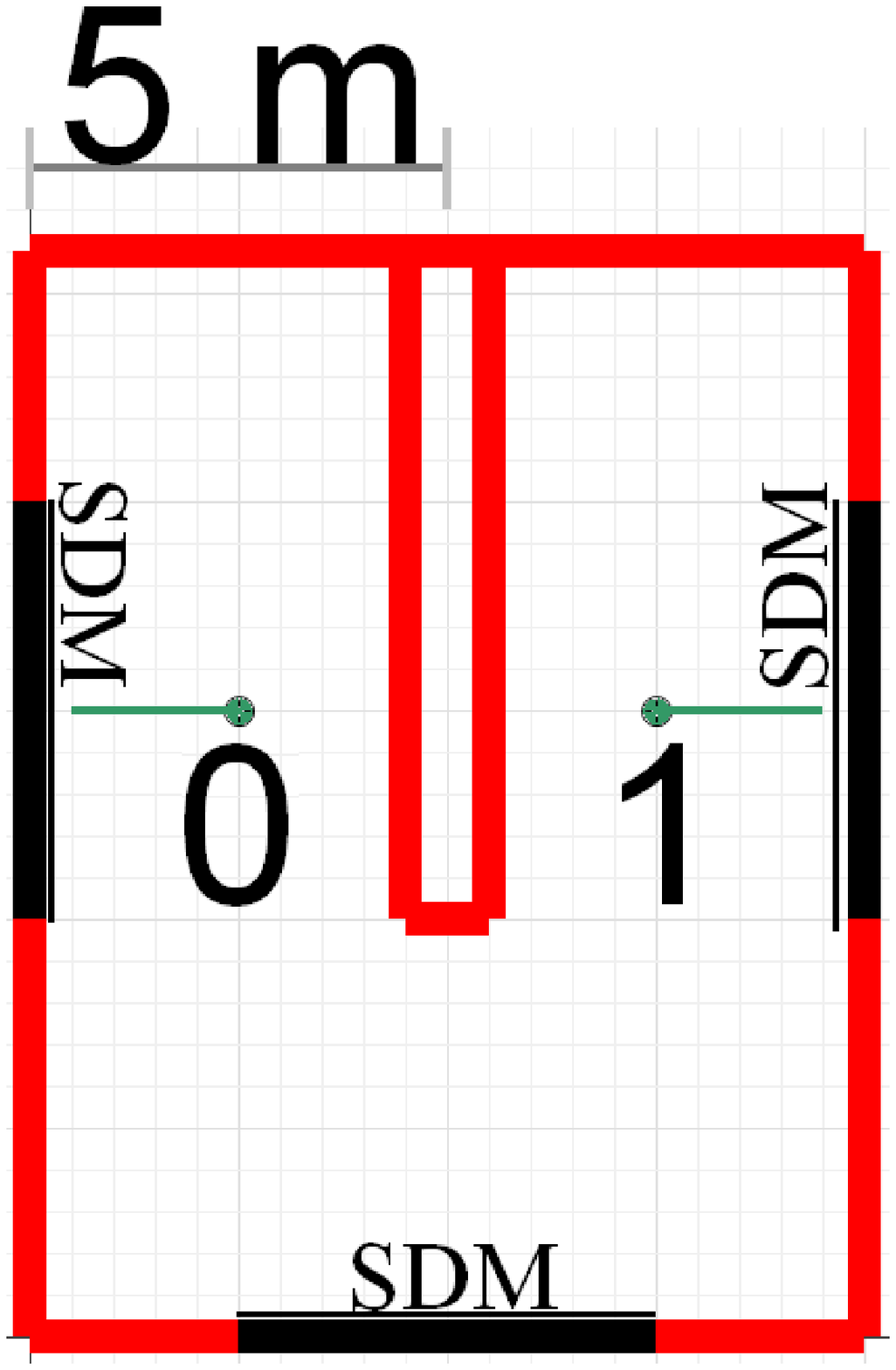}\includegraphics[width=0.24\columnwidth]{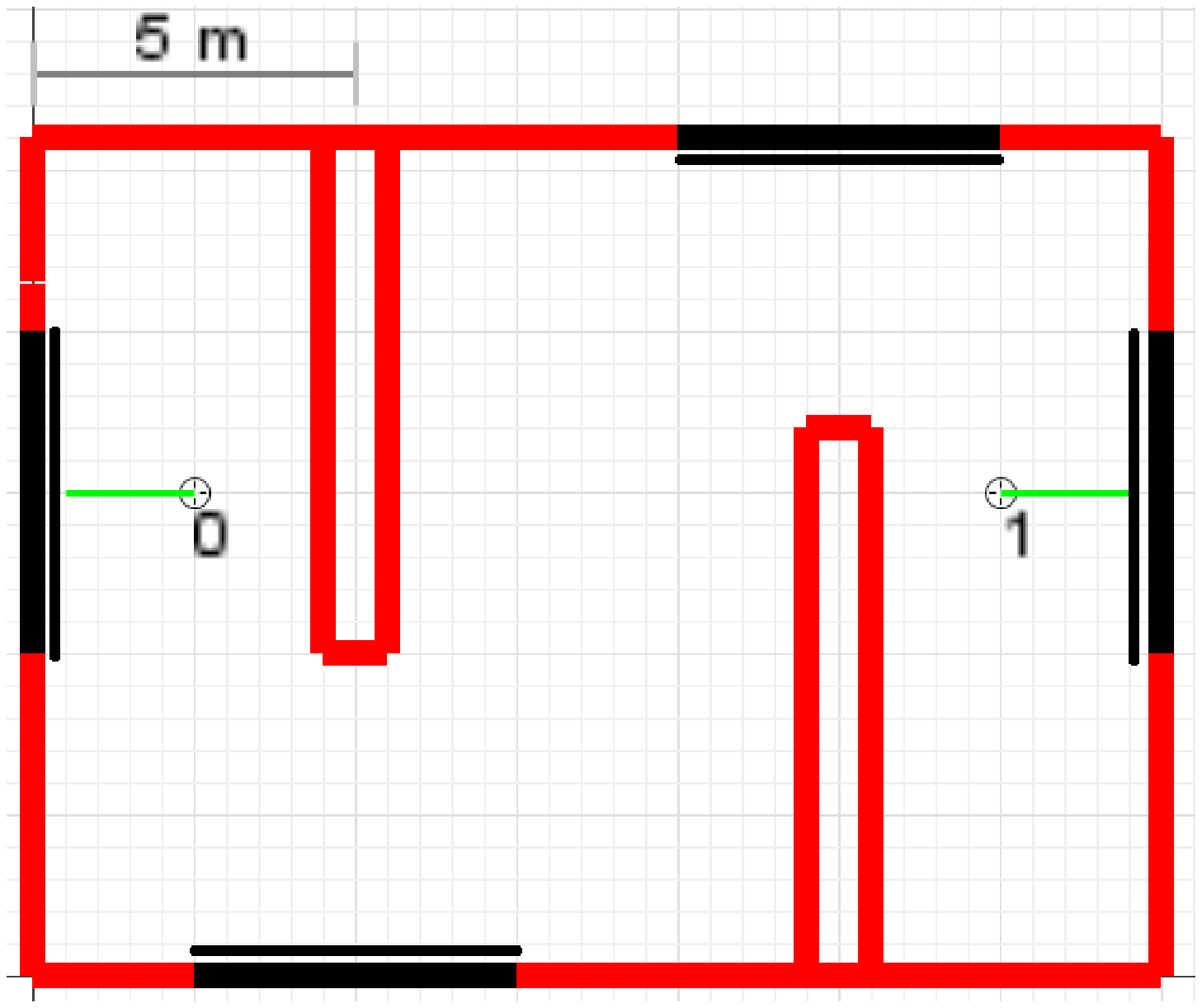}\includegraphics[width=0.35\columnwidth]{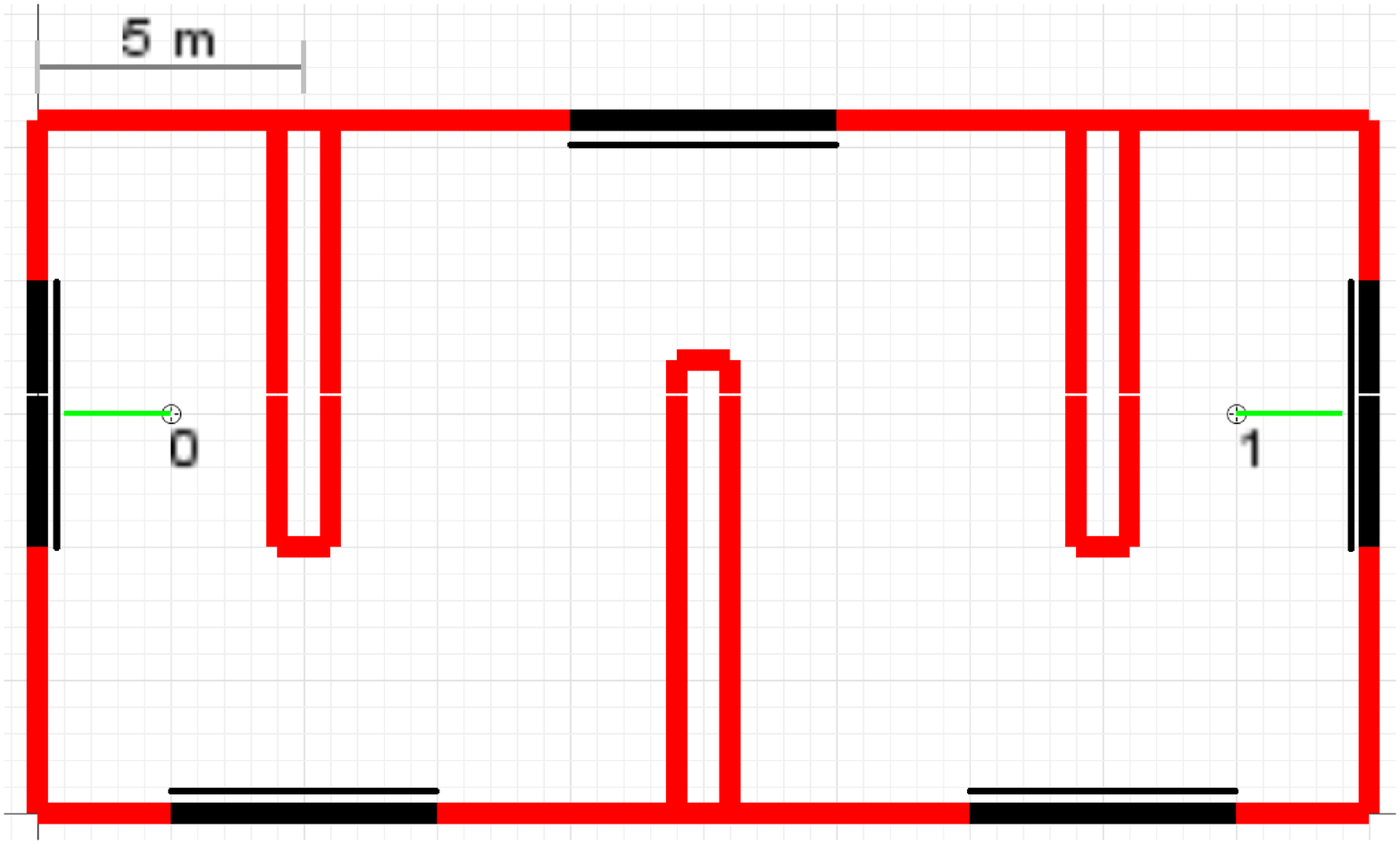}}\tabularnewline
\multicolumn{2}{c}{\includegraphics[width=0.39\columnwidth]{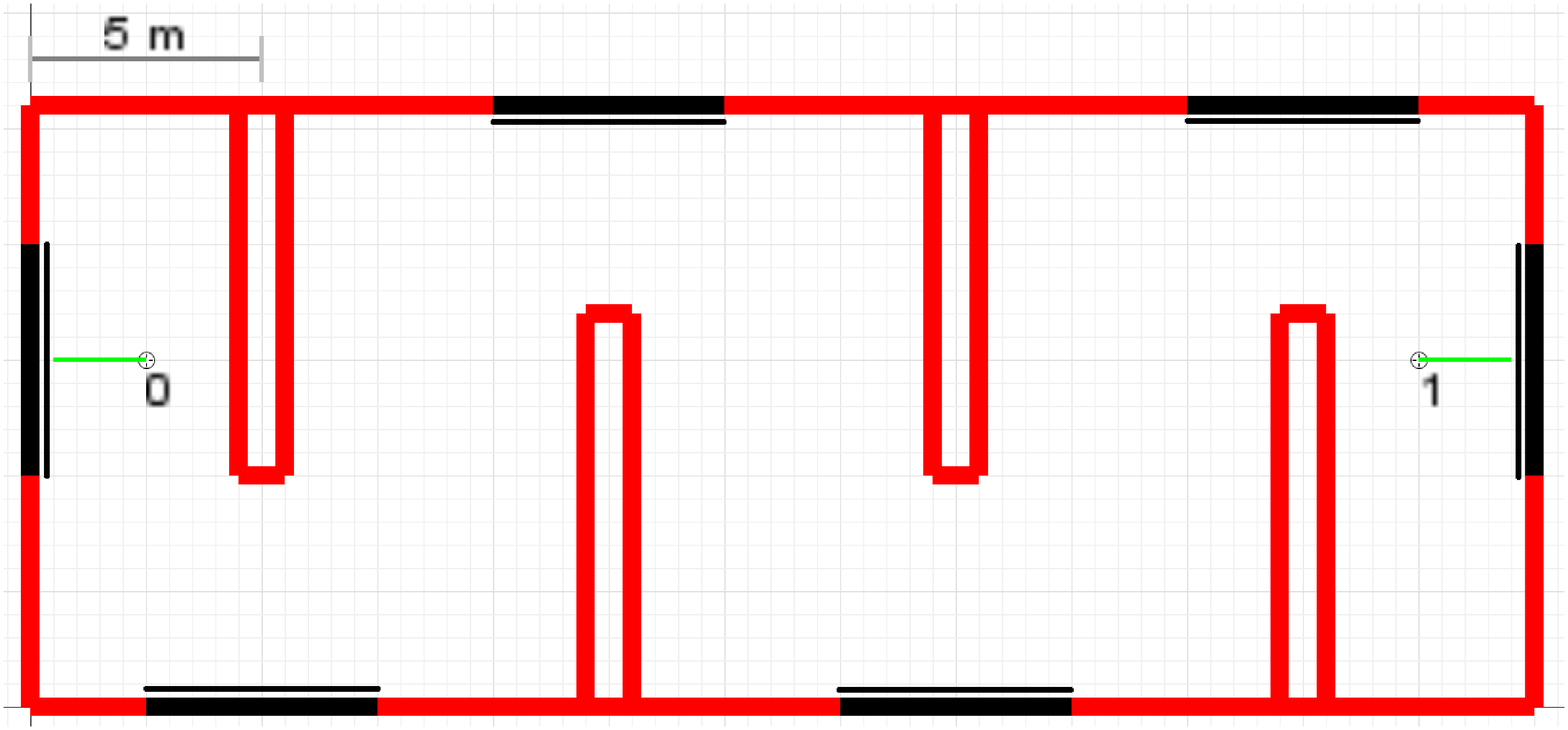}\includegraphics[width=0.45\columnwidth]{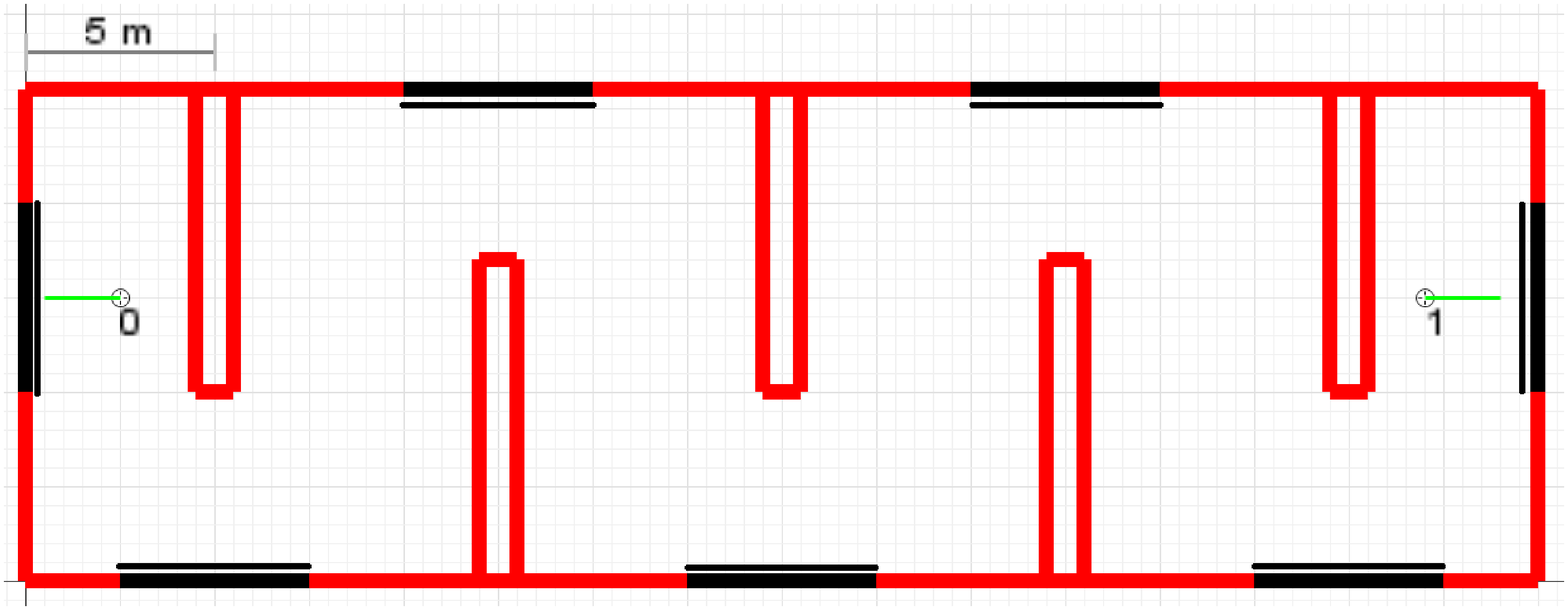}}\tabularnewline
\multicolumn{2}{c}{%
\begin{tabular}{@{}c@{}}
Five floorplans studied \tabularnewline
(\textcolor{black}{with increasing number of middle-walls from one
to five}).\tabularnewline
\end{tabular}}\tabularnewline
\end{tabular}
\end{table}

\begin{table}[t]
\centering{}{\small{}\caption{\textsc{\label{tab:TSimParams}Simulation parameters.}}
}%
\begin{tabular}{|c|c|}
\hline
{\small{}Ceiling Height } & {\small{}$3$~$m$}\tabularnewline
\hline
{\small{}Tile Dimensions } & \textbf{\small{}$1\times1\,m$}{\small{} (z-centered)}\tabularnewline
\hline
{\small{}Tile Functions } & {\small{}$\text{\textsc{Steer}}$, $\text{\textsc{Split}}$, $\text{\textsc{Focus}}$,
$\text{\textsc{Absorb}}$}\tabularnewline
\hline
{\small{}Non-SDM surfaces } & {\small{}}%
\begin{tabular}{@{}c@{}}
{\small{}Perfect absorbers }\tabularnewline
{\small{}(cf. red in Table~\ref{tab:Floorplan})}\tabularnewline
\end{tabular}\tabularnewline
\hline
{\small{}Frequency } & {\small{}$2.4\,GHz$ }\tabularnewline
\hline
{\small{}TX Power } & {\small{}$-30\,dBm$}\tabularnewline
\hline
{\small{}Antenna type } & {\small{}Single $a^{o}$-lobe sinusoid,}\tabularnewline
 & {\small{}pointing at $\phi^{o},\theta^{o}$ (cf. Table~\ref{tab:Floorplan})}\tabularnewline
\hline
{\small{}Max ray bounces } & {\small{}$50$}\tabularnewline
\hline
{\small{}Power loss per bounce } & {\small{}$1$~\%}\tabularnewline
\hline
{\small{}NN training cycles } & {\small{}$10,000$}\tabularnewline
\hline
\textcolor{black}{\small{}NN optimization} & \textcolor{black}{\small{}Gradient with momentum}\tabularnewline
\hline
{\small{}NN pruning factor range } & {\small{}20\%~:~20\%~:~100\%}\tabularnewline
\hline
\end{tabular}
\end{table}

In each floorplan we consider a TX-RX pair as shown in Table~\ref{tab:Floorplan}.
For ease of exposition, i.e., in order to provide meaningful illustrations
of the actual wireless propagation, SDM-covered surfaces are denoted
as black-colored walls, thereby confining the tile-sets produced by
\noun{KpConfig} into these locations. We seek to tune each SDM tile
in this setup so as to maximize the received power at the RX. It is
noted that, based on the \noun{KpConfig} workflow, in a multi-user
scenario each pair is treated sequentially, i.e., walls are defined
by \noun{KpConfig}, then one neural network is created and trained
by \noun{NNConfig}, finally configuring the corresponding tiles. The
process repeats for each pair, each time considering any previously
configured tiles as frozen, i.e., subject to no further training in
the neural network sense. Thus, the single-pair scenario performance
reflects the core, repeating process which is also coupled with ease
of exposition benefits.
\begin{figure}[tp]
\begin{centering}
\subfloat[\label{fig:KpPropag1wall}\noun{KpConfig} propagation.]{\begin{centering}
\includegraphics[width=0.4\columnwidth]{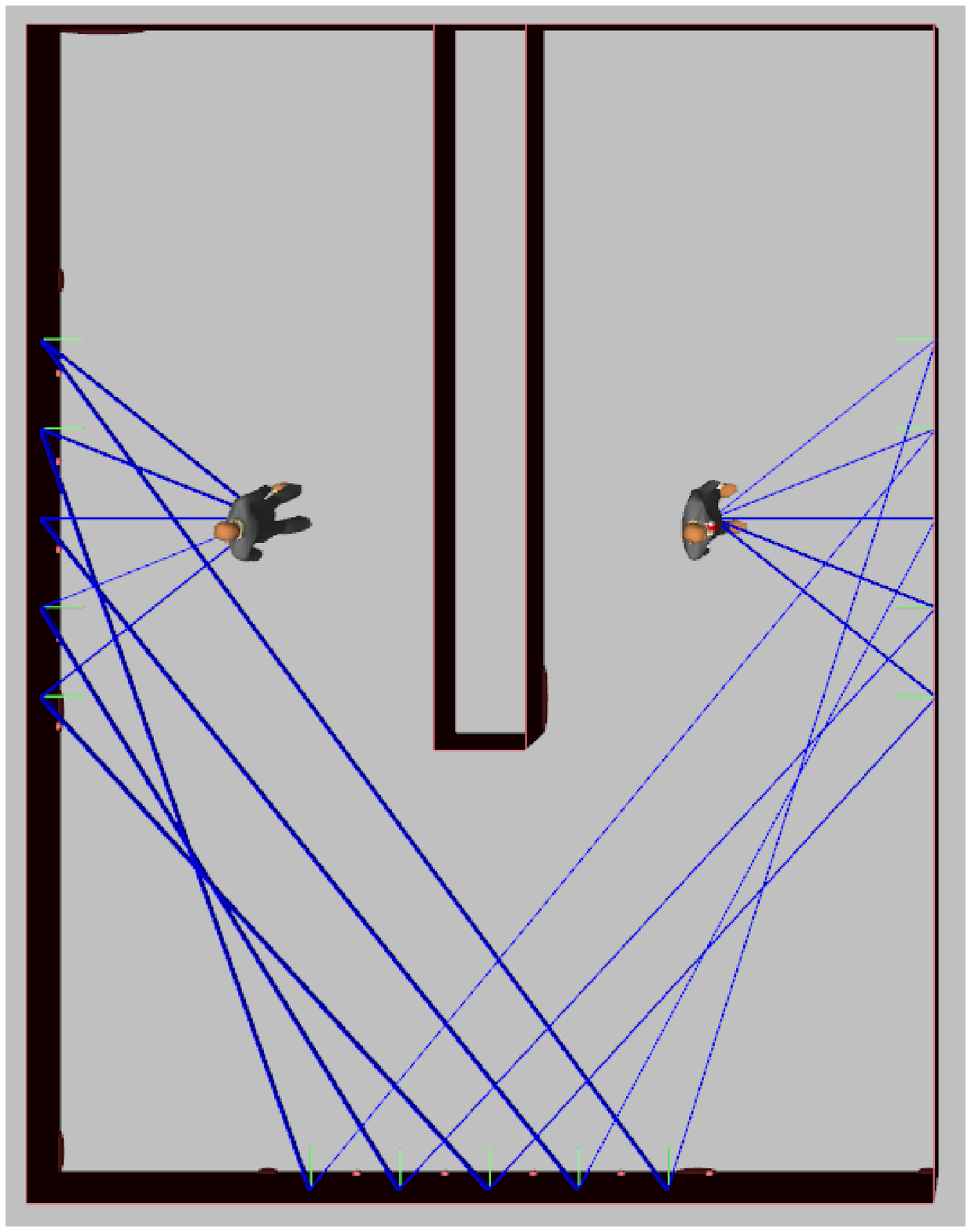}
\par\end{centering}
}\subfloat[\label{fig:NnPropag1wall}\noun{NNConfig} propagation.]{\begin{centering}
\includegraphics[width=0.4\columnwidth]{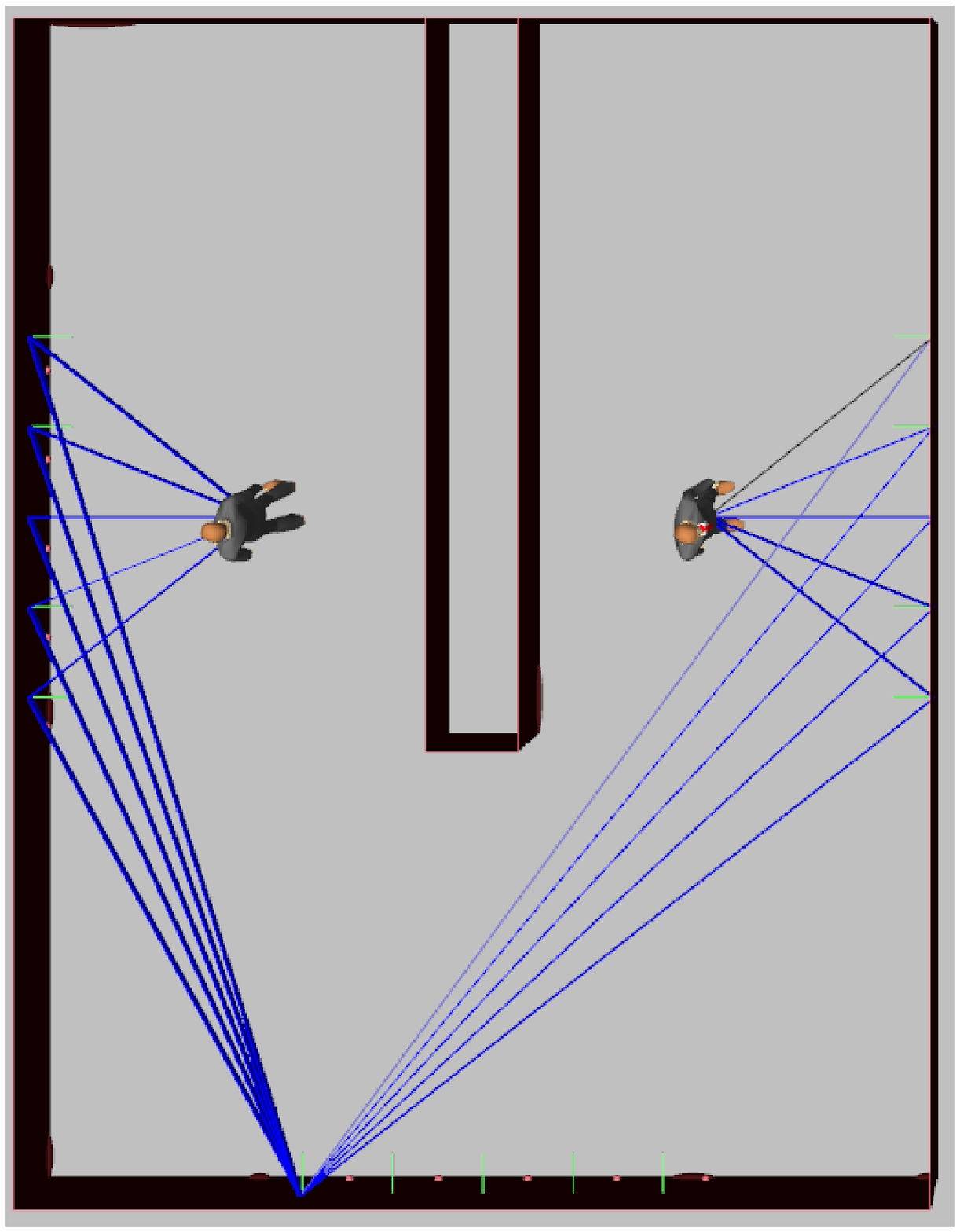}
\par\end{centering}
}
\par\end{centering}
\begin{centering}
\subfloat[\label{fig:NeuralNet1wall}Trained neural network corresponding to
Fig.~\ref{fig:NnPropag1wall}.]{\begin{centering}
\includegraphics[width=0.75\columnwidth]{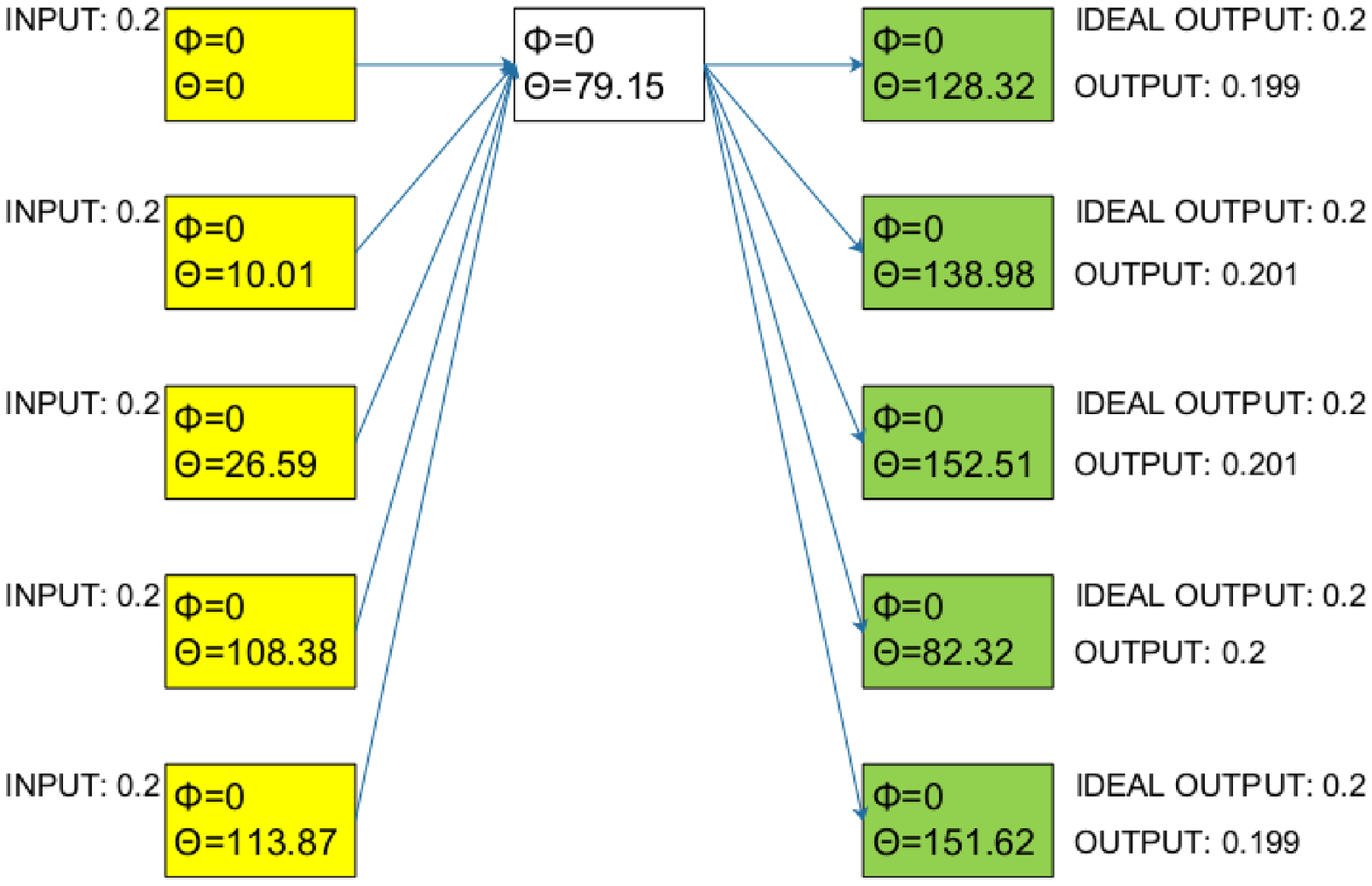}
\par\end{centering}
}
\par\end{centering}
\caption{\label{fig:Propag1wall}KpConfig and NNConfig behavior at the 1-middle
wall floorplan case. In KpConfig-derived propagation, individual paths
fully occupy available tiles, whereas in NNConfig-derived propagation
only the single-best tile is selected which improves received power
and link efficiency.}
\end{figure}

Table~\ref{tab:TSimParams} summarizes the persistent simulator parameters
across all subsequent tests. As in~\cite{liaskos2019network}, each
\noun{Steer} or \noun{Split} function is coupled with a wave \noun{Focus}
function at the same direction. \noun{Focus} essentially counter-balances
the free-space path loss by a corresponding reflection gain, which
is a unique capability of metasurfaces over reflectarrays \textcolor{black}{and
phased array} antennas. Thus, the overall path loss is essentially
defined by any power consumed over the metasurface materials per each
bounce~\cite{liaskos2019network}. This loss has been shown to take
very small values, at the order of 1\%~\cite{MSSurveyAllFunctionsAndTypes}.
\textcolor{black}{The NN pruning factor is an input variable (common
for all neural network layers), whose studies range is 20\% to 100\%
with steps of 20\%. }

Once the neural network is created, the inputs and ideal outputs are
set to their values, as described in Section~\ref{sec:ensembleNN}.
We employ the root mean square error (RMSE) as the deviation $\xi$
between the ideal outputs and the training outcomes in each cycle.
The initial $\theta,\,\phi$ angle values (azimuth and elevation of
the virtual normal) per neural node are randomized in the ranges of~$\left[-90^{o},90^{o}\right]$
and~$\left[0^{o},90^{o}\right]$, respectively. Finally, the termination
criterion is to reach $10,000$ feed-forward/back-propagate cycles,
which is kept constant over all floorplan cases to ensure a constant
runtime.
\begin{figure}[tp]
\begin{centering}
\subfloat[\label{fig:KpPropag5wall}\noun{KpConfig} propagation.]{\begin{centering}
\includegraphics[width=0.9\columnwidth]{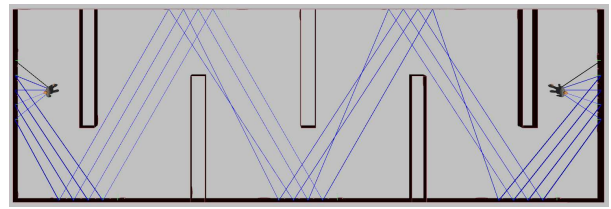}
\par\end{centering}
}
\par\end{centering}
\begin{centering}
\subfloat[\label{fig:NnPropag5wall}\noun{NNConfig} propagation.]{\begin{centering}
\includegraphics[width=0.9\columnwidth]{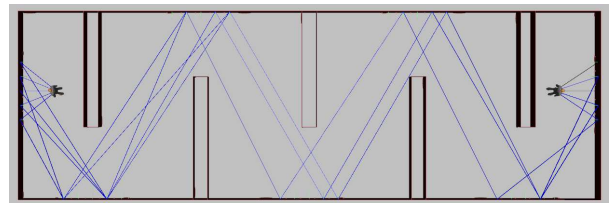}
\par\end{centering}
}
\par\end{centering}
\caption{\label{fig:Propag5walls}KpConfig and NNConfig behavior at the 5-middle
wall scenario. Such comparison in the number of deployed tiles demonstrates
NNConfig's advantage on link efficiency enhancement.}
\end{figure}

Figs.~\ref{fig:Propag1wall}, \ref{fig:Propag2walls}, and~\ref{fig:Propag5walls}
visualize the end-to-end propagation achieved by the proposed \noun{NNConfig}
and the \noun{KpConfig} over the floorplans 1, 2, and 5, as identified
by the number of middle walls. (Floorplans 3 and 4 convey similar
conclusions and are omitted).

In the floorplan 1 case (Fig.~\ref{fig:Propag1wall}), the \noun{KpConfig}
outcome (Fig.~\ref{fig:KpPropag1wall}) visualizes the disadvantage
of this approach detailed in Section~\ref{sec:Related-work}: each
user link is treated sequentially and connected to its end-point within
a graph, resulting into a multitude of independent paths and full
occupancy of all available tiles. On the other hand, \noun{NNConfig}
achieves in finding a propagation solution that employs just a single
tile, with a \textcolor{black}{\noun{MultiSteer}}{\small{} }\textcolor{black}{functionality
(Algorithm~\ref{alg:multiSteer}). Single }\textcolor{black}{\noun{Steer}}\textcolor{black}{{}
functionalities are necessarily applied to the tiles adjacent to the
users in both schemes. The trained neural network is shown in Fig.~\ref{fig:NeuralNet1wall},
which shows good correspondence with the mapped result of Fig.~\ref{fig:NnPropag1wall}.
 }

In the floorplan 2 case (Fig.~\ref{fig:Propag2walls}), \noun{KpConfig}
once again ensures connectivity by using up all available tiles~(Fig.~\ref{fig:KpPropag2wall}).
\noun{NNConfig} manages to also achieve full connectivity while using
5 tiles less, by employing 4 \noun{MultiSteer} functionalities. In
the floorplan 5 case (Fig.~\ref{fig:Propag5walls}), which potentially
constitutes an extreme scenario, \noun{KpConfig} exhibits the same
behavior, i.e., full connectivity and full tile occupancy (Fig.~\ref{fig:KpPropag5wall}).
\noun{NNConfig} pinpoints several \noun{MultiSteer} functionalities,\textcolor{black}{{}
combining the transmitting user's emissi}ons over much fewer tiles
towards the receiver~(Fig.~\ref{fig:NnPropag5wall}).\textcolor{black}{{}
It is worth noting that the air-paths demonstrate a pattern of ``two-cluster''
formation in Figs.~\ref{fig:NnPropag2wall}~and~\ref{fig:NnPropag5wall},
the reason for such coincidence is due to the symmetries of the two
floorplans and is not a deterministic input to the }\textcolor{black}{\noun{NNConfig}}\textcolor{black}{.
}
\begin{figure}[tp]
\begin{centering}
\includegraphics[clip,width=1\columnwidth,viewport=0bp 35bp 960bp 500bp]{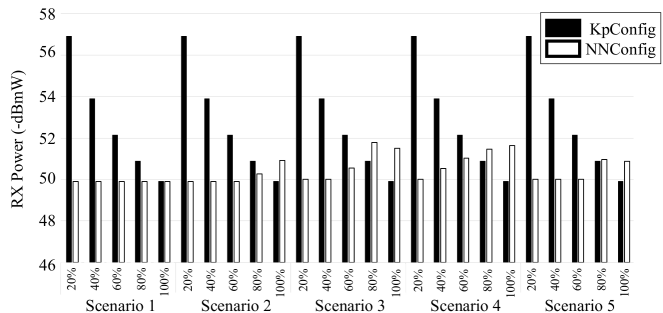}
\par\end{centering}
\caption{\textcolor{blue}{\label{fig:RxPow-1}}Received power per compared
scheme over the studied floorplans. The y-axis is in negative $dBmW$
(i.e., lower is better). The $20-100%\%
$ percentage represents the neural network pruning factor.}
\end{figure}

\textcolor{black}{The absolute received power performance of the compared
schemes is shown in Fig.~\ref{fig:RxPow-1}. Notably, the performance
of }\textcolor{black}{\noun{KpConfig}}\textcolor{black}{{} depends strongly
on the the neural network pruning factor: higher factor values (i.e.,
neural network layers containing }\textcolor{black}{\emph{more}}\textcolor{black}{{}
nodes) favor the }\textcolor{black}{\noun{KpConfig}}\textcolor{black}{{}
performance, and is repeated across all floorplan scenarios. This
is a natural outcome, given that }\textcolor{black}{\noun{KpConfig}}\textcolor{black}{{}
does not reuse tiles, i.e., one tile serves exactly one impinging
wave direction, as shown in Fig.~\ref{fig:KpPropag1wall}. Removing
a tile leads to a proportional reduction in the successfully steered
(and RX-received) wireless power. }\textcolor{black}{\noun{NNConfig}}\textcolor{black}{{}
exhibits weaker dependence from the pruning factor, and received power
near the }\textcolor{black}{\noun{KpConfig}}\textcolor{black}{{} best
case. In fact, }\textcolor{black}{\noun{NNConfig}}\textcolor{black}{{}
benefits from low factor values (i.e., neural network layers containing
}\textcolor{black}{\emph{less}}\textcolor{black}{{} nodes), since the
algorithm is pushed to: i) find PWE configurations with increased
tile reuse, and ii) a smaller neural network is likely to be trained
better in allotted time. It is noted that the received power in the
non-PWE case (no SDMs) is below $-150\,\text{dBmW}$ in all floorplans.}
\begin{figure}[tp]
\begin{centering}
\includegraphics[clip,width=1\columnwidth,viewport=0bp 50bp 960bp 520bp]{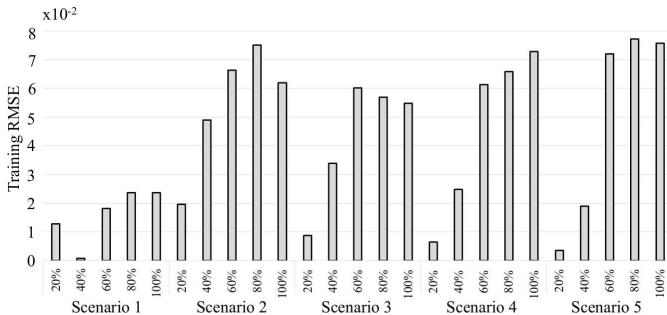}
\par\end{centering}
\caption{\textcolor{blue}{\label{fig:RMSE}}Training Root Mean Square Error
for NNConfig over the studied floorplans.}
\end{figure}

\textcolor{black}{To better understand the remarked behavior of }\textcolor{black}{\noun{NNConfig}}\textcolor{black}{{}
we turn to Fig.~}\ref{fig:RMSE}. Setting a static limit to the neural
network training cycles (cf. Table~\ref{tab:TSimParams}) yields
an increase in the training RMSE, given the increasing network size
per floorplan. However, this error is almost inconsequential to the
overall performance, as was shown in the context of Fig.~\ref{fig:RxPow-1}.
This is due to the fact that the neural network interpretation process
is forgiving to link weight imprecisions owed to prematurely stopped
training. In essence, the interpretation process of Section~\ref{subsec:Interpretability:-Mapping-a}
revolves around node connectivity, rather than precise link weights.
Thus, if the main neural paths have been established, the interpretation
process will yield a good outcome, even if the link weights have not
been fully optimized.
\begin{figure}[p]
\begin{centering}
\subfloat[\textcolor{blue}{\label{fig:MOMindependence}}Momentum independence
(learning rate is $1.0$).]{\begin{centering}
\includegraphics[clip,width=1\columnwidth,viewport=0bp 60bp 960bp 505bp]{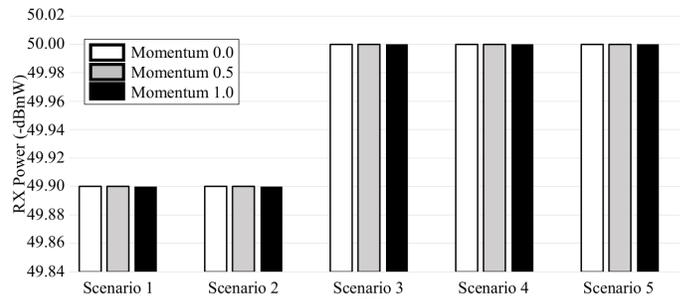}
\par\end{centering}
}
\par\end{centering}
\begin{centering}
\subfloat[\textcolor{blue}{\label{fig:LERindependene}}Learning rate independence
(momentum is $1.0$).]{\begin{centering}
\includegraphics[clip,width=1\columnwidth,viewport=0bp 60bp 960bp 510bp]{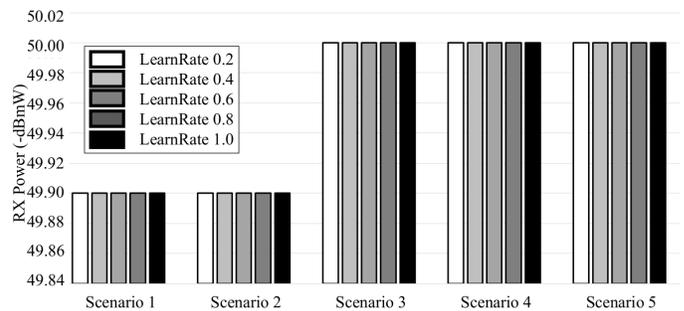}
\par\end{centering}
}
\par\end{centering}
\caption{\label{fig:independence}Validation of the independence of the NNConfig
performance from the classic neural network training parameters (pruning
factor is $20$\%).}
\end{figure}

Regarding the neural network training process, as discussed in the
Annex, we employ a standard feedforward/backpropagation operation~\cite{islam2001new},
but alter the link weight deltas to comply with relation~(\ref{eq:weightalter})
at each step. In the general case, (i.e., not specialized to NNConfig)
a backpropagation process updates the link weights at a given $step$
as:
\begin{equation}
w_{ji}'\gets w_{ji}-\eta\cdot\left(\frac{\partial\xi}{\partial w_{ji}}\right)_{step}-\mu\cdot\left(\frac{\partial\xi}{\partial w_{ji}}\right)_{step-1}\label{eq:general}
\end{equation}
where $\eta$ is a learning rate factor and $\mu$ is the momentum
factor attributed to the partial derivative for link $ji$ at the
previous backpropagation step $t-1$, and then we alter the $w_{ji}'$
. For the NNConfig-specialized process, we subsequently alter each
$w_{ij}'$ by a multiplicative factor $\eta_{ji}$, to ensure compliance
with relation~(\ref{eq:weightalter}). This raises the question:
do we need to optimize the factors $\eta$ and $\mu$ of the general
update rule~(\ref{eq:general}) for NNConfig (i.e., prior to weight
adaptation)? In Fig.~\ref{fig:independence} we execute a parametric
variation study of the received power for NNConfig in all five floorplan
scenarios. Notably, the NNConfig performance exhibits no dependence
on the factors $\eta$ and $\mu$. The required compliance with relation~(\ref{eq:weightalter})
essentially defines any multiplicative factor for the quantity $\frac{\partial\xi}{\partial w_{ji}}$,
meaning that NNConfig does not need to be tuned in that aspect, discarding
the dependence from two of the most common neural network parameters
(learning rate and momentum) in general and limiting the size of its
solution space.
\begin{figure}[tp]
\begin{centering}
\includegraphics[viewport=0bp 110bp 960bp 450bp,clip,width=1\columnwidth]{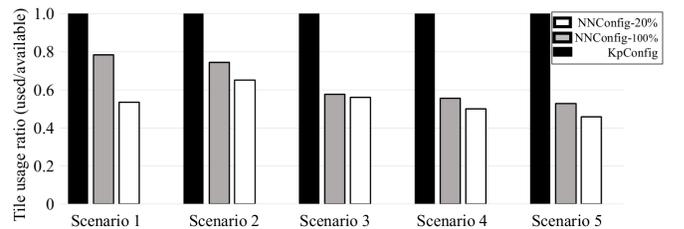}
\par\end{centering}
\caption{\label{fig:TILES-1}Ratio of SDM tiles used per compared scheme over
the studied floorplans.}
\end{figure}

Limiting the tile occupancy ratio is important to the operation of
a PWE for two reasons. First, it implies that the same PWE can serve
more communicating user pairs, thereby yielding higher served user
capacity~\cite{liaskos2019network}. Second, tiles are electronic
devices that need to be powered and be in communication with the PWE
server, resulting into electrical power consumption, as well as into
computational and communication system overhead. Therefore, limiting
their use is beneficial from these aspects\textcolor{black}{.}\textcolor{black}{\noun{
NNConfig}}\textcolor{black}{{} presents considerable gains in}\textcolor{black}{\noun{
}}\textcolor{black}{number of tiles occupied, as shown in Fig.~\ref{fig:TILES-1}.
In this Figure we compare }\textcolor{black}{\noun{KpConfig}}\textcolor{black}{{}
(pruning factor 100\%) and }\textcolor{black}{\noun{NNConfig}}\textcolor{black}{{}
(pruning factor 20\%) variations, since they yield the maximal received
power for both algorithms (cf. Fig.~\ref{fig:RxPow-1}). Moreover,
we include the }\textcolor{black}{\noun{NNConfig}}\textcolor{black}{{}
(pruning factor 100\%) to the comparison, to demonstrate the tile
number-economic operation of }\textcolor{black}{\noun{NNConfig}}\textcolor{black}{{}
even in the worst case. }\textcolor{black}{\noun{NNConfig}}\textcolor{black}{-100\%
uses 75\% of the available tiles in the floorplan~1 case, a percentage
that reduces to approximately 50\% in the floorplan~5 scenario. In
contrast, the }\textcolor{black}{\noun{KpConfig}}\textcolor{black}{{}
tile occupancy is always 100\%. In essence, each scheme necessarily
uses all tiles in the walls adjacent to the users (as these receive
the users' emissions), but differ in the use of tiles in the intermediate
surfaces. Moreover, each floorplan naturally introduces more available
tiles, due to the enlargement of the environment, as shown in Table~\ref{tab:Floorplan}.
Nonetheless, the }\textcolor{black}{\noun{NNConfig}}\textcolor{black}{{}
tile occupancy is low, at a level that the overall trend of the plot
in Fig.~\ref{fig:TILES-1} is to decrease. }\textcolor{black}{\noun{NNConfig-20\%}}\textcolor{black}{{}
performs better overall, yielding the highest received power in Fig.~\ref{fig:RxPow-1}.
}
\begin{figure}
\begin{centering}
\subfloat[\label{fig:PrxVar}Received power variation.]{\begin{centering}
\includegraphics[viewport=0bp 80bp 960bp 450bp,clip,width=0.95\columnwidth]{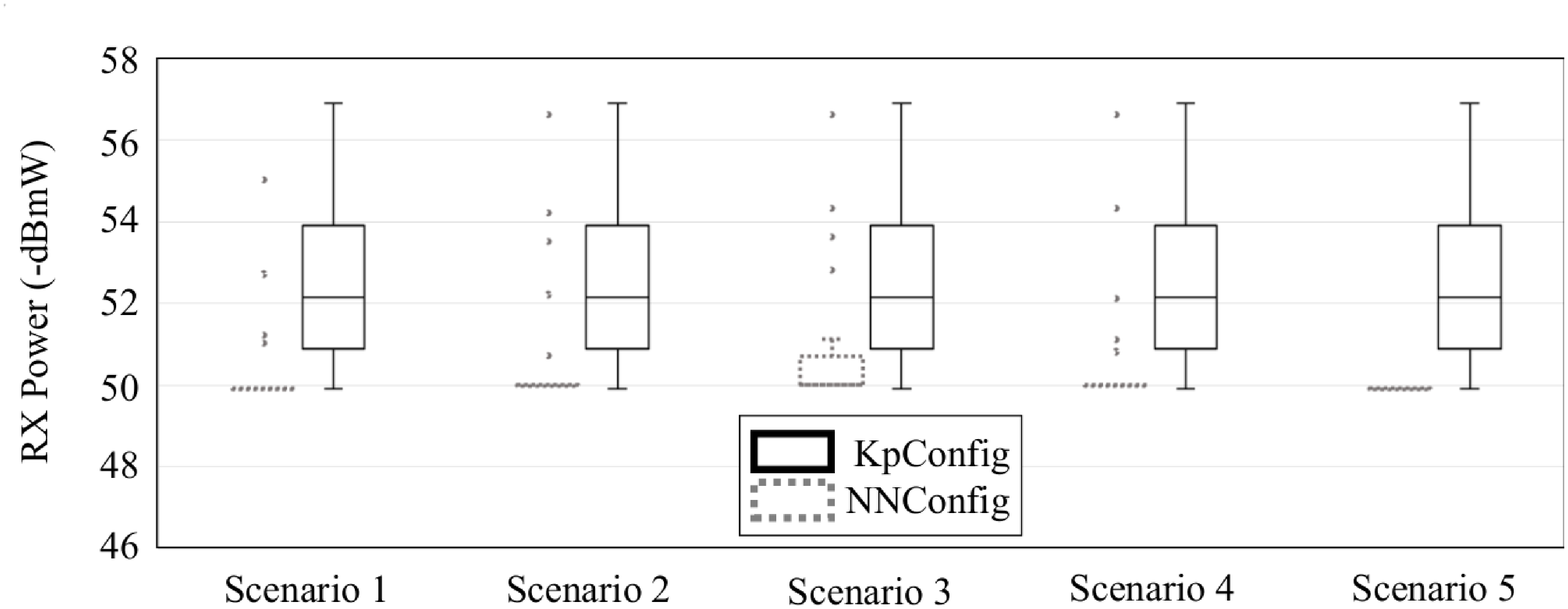}
\par\end{centering}
}
\par\end{centering}
\begin{centering}
\subfloat[\label{fig:TILEvar}Tile usage variation.]{\begin{centering}
\includegraphics[width=0.95\columnwidth,viewport=0bp 0bp 1342bp 482bp]{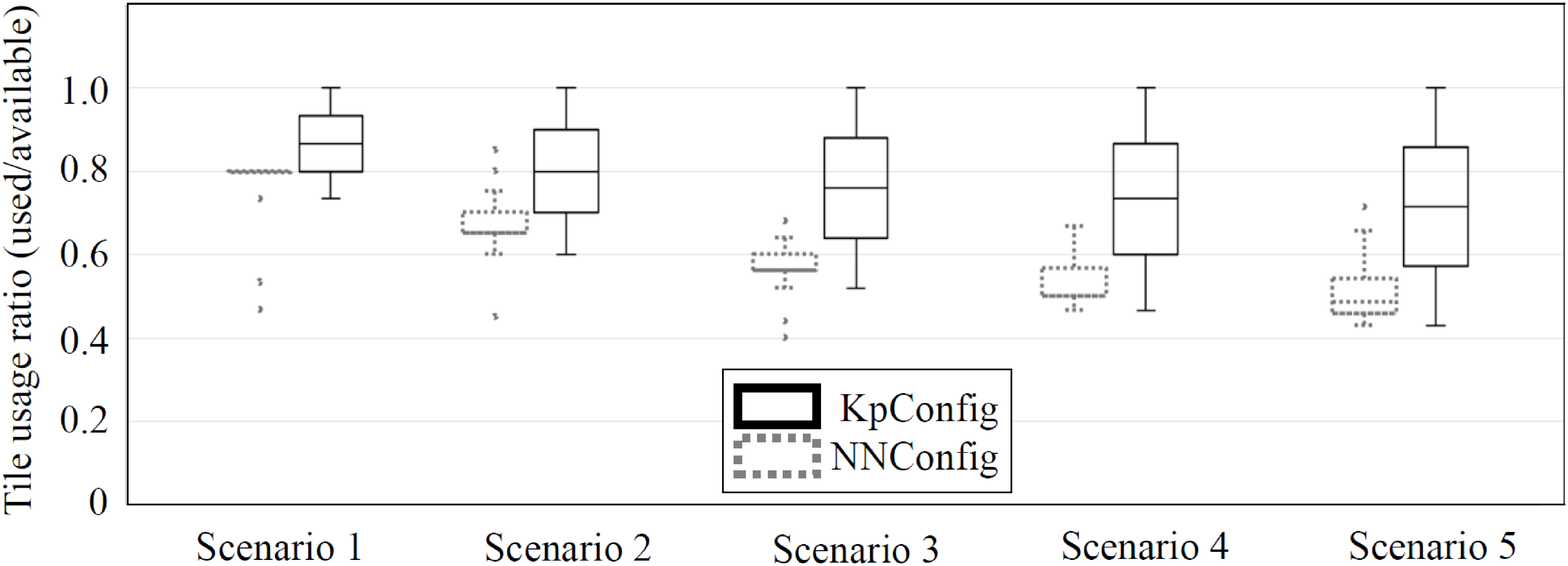}
\par\end{centering}
}
\par\end{centering}
\caption{\label{fig:boxplot}Performance variation of\noun{ NNConfig} and\noun{
KpConfig} over the complete range of pruning factor values (20\%-100\%
step by 20\%).}
\end{figure}

\textcolor{black}{In the preceding experiments it was made evident
that the pruning factor is the only input parameter of }\textcolor{black}{\noun{NNConfig}}\textcolor{black}{{}
which is potentially subject to optimization. Therefore, in Fig.~\ref{fig:boxplot}
we study the performance variation of }\textcolor{black}{\noun{NNConfig}}\textcolor{black}{{}
and }\textcolor{black}{\noun{KpConfig}}\textcolor{black}{{} over the
complete range of the pruning factor given in Table~\ref{tab:TSimParams}.
Our goal is to evaluate the performance bounds assuming no means of
optimizing the pruning factor. As shown in Fig.~\ref{fig:PrxVar},
}\textcolor{black}{\noun{NNConfig}}\textcolor{black}{{} exhibits limited
RX power dependence: regardless in the number of available tiles is
large, }\textcolor{black}{\noun{NNConfig}}\textcolor{black}{{} will
search for tile-reusing PWE configurations. Therefore, assuming enough
runtime, the solution will be consistent and invariable. On the other
hand,}\textcolor{black}{\noun{ KpConfig}}\textcolor{black}{{} is very
sensitive to the pruning factor, as it is greedy in tile usage and
the exclusion of tiles from the solution space can reduce the received
power proportionately (cf.~Fig.~\ref{fig:KpPropag1wall} and Fig.~\ref{fig:RxPow-1}).
Thus, the }\textcolor{black}{\noun{KpConfig}}\textcolor{black}{{} variation
is large and uniform across all floorplan scenarios. The tile usage
variation in Fig.~\ref{fig:TILEvar} follows the same rationale,
and the trend is similar to Fig.~\ref{fig:TILES-1}.}
\begin{figure}[tp]
\begin{centering}
\includegraphics[viewport=0bp 0bp 415bp 170bp,width=1\columnwidth]{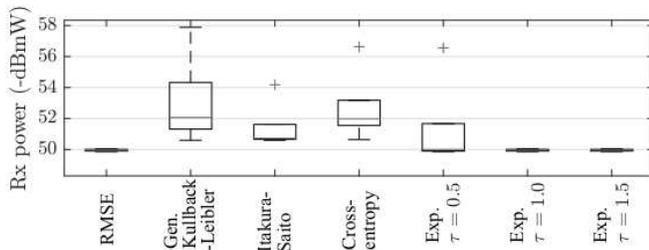}
\par\end{centering}
\caption{\label{fig:CostFunct}\textcolor{black}{Performance variation of}\textcolor{black}{\noun{
NNConfig}}\textcolor{black}{{} for various cost functions.}}
\end{figure}

\textcolor{black}{Finally, in Fig.~\ref{fig:CostFunct} we justify
the use of RMSE as the cost function. After experimentation with several
alternatives (listed in the x-axis of Fig.~\ref{fig:CostFunct}),
RMSE and the exponential cost function, $\tau\cdot e^{\sum_{\forall l}\nicefrac{1}{\tau}\left(\delta_{l}\right)^{2}}$,
behaved equally well in terms of yielding the best performance over
all experiments and layout scenarios. Nonetheless, the exponential
cost function requires the optimization of its tunable parameter $\tau$,
while RMSE does not pose such a requirement. }

\section{Conclusion\label{sec:Conclusion}}

Intelligent surfaces allow for programmatic control over the wireless
propagation phenomenon within a space, with novel capabilities in
wireless link quality, security and power transfer. The largest the
deployment of surface units within an indoor or outdoor space, the
more deterministic the control over the wireless propagation as a
whole, to the benefit of communicating users. The present study proposed
a novel technique for such large scale configurations based on machine
learning techniques. Particularly, a customized, interpretable neural
network was proposed, where surface units are represented as neural
network nodes, and connectivity as neural links. The neural network
is required to yield ideal output, such as total signal delivery to
the receiver, and back-propagation rules optimize the link weights
accordingly. An interpretation process maps the trained links to intelligent
surface configurations. \textcolor{black}{The proposed scheme inherits
the weaknesses of neural networks deriving from their heuristic nature:
convergence guarantees within a given time interval cannot be given.
As such, the scheme outputs should always be compared to simpler but
deterministic solutions, and be adopted when clearly superior. Simulations
showed that the proposed scheme can be synergistic to related schemes,
and effectively use a single surface towards meeting multiple objectives,
thus being economic in terms of total surfaces used for serving a
set of users.}

\section*{Acknowledgment}

This work was funded by the European Union via grants EU736876, and
EU833828.

% Generated by IEEEtran.bst, version: 1.14 (2015/08/26)

\appendix
\textcolor{black}{\label{sec:appendix} Assume a neural network with
the following notation: i) $x_{ji}$\textendash{} the $i^{th}$ input
to node $j$, ii) $w_{ji}$\textendash{} the weight associated with
input $i$ to unit $j$, iii) $z_{j}$\textendash{} the weighted sum
of inputs for node $j$ ($z_{j}=\sum_{\forall i}w_{ji}\cdot x_{ji}$
). }

\textcolor{black}{The linear ramp function will serve as the activation
function and, thus, $z_{j}$ will also be equal to the output of node
$j$. This simplifies the classic neural network derivatives as~\cite{islam2001new}:
\begin{equation}
\frac{\partial\xi}{\partial w_{ji}}=\frac{\partial\xi}{\partial z_{j}}\cdot\frac{\partial z_{j}}{\partial w_{ji}}=\frac{\partial\xi}{\partial z_{j}}\cdot x_{ji},\label{eq:generic}
\end{equation}
which holds in general. Additional simplifications are:
\begin{equation}
\frac{\partial\xi}{\partial z_{j}}=-\left(o_{j}-z_{j}\right)\label{eq:output}
\end{equation}
for any output layer node and $\xi:\,RMSE$. For a hidden layer node
$j$, we use the notation $\left\{ \underset{\to}{j}\right\} $ to
denote all nodes downstream of $j$ and indexed by $k$. These nodes
are affected by the output of node $j$ and, therefore:
\begin{equation}
\frac{\partial\xi}{\partial z_{j}}=\sum_{\forall k\in\left\{ \underset{\to}{j}\right\} }\frac{\partial\xi}{\partial z_{k}}\cdot\frac{\partial z_{k}}{\partial z_{j}}=\sum_{\forall k\in\left\{ \underset{\to}{j}\right\} }\frac{\partial\xi}{\partial z_{k}}\cdot w_{kj}\label{eq:hiddenD}
\end{equation}
Finally, for any node (output or hidden) we note the well-known weight
update rule:
\begin{equation}
w_{ji}^{*}\gets w_{ji}-\eta\cdot\frac{\partial\xi}{\partial w_{ji}}\label{eq:updateGeneric}
\end{equation}
where $\eta$ is a constant (commonly referred to as learning rate).
$\frac{\partial\xi}{\partial w_{ji}}$ is derived from eq.~(\ref{eq:generic})
and (\ref{eq:hiddenD}) for hidden nodes, and via eq.~(\ref{eq:generic})
and (\ref{eq:output}) for output nodes. Up to this point, the solution
space of $w_{ij}$ values is completely unrestricted: $w_{ij}$ values
are completely disjoint.}

\textcolor{black}{Specializing for the neural networks studies in
this paper, we derive from eq.~(\ref{eq:weights-1}): }
\begin{itemize}
\item \textcolor{black}{The weight values sourced from a node (downstream)
need to be greater than zero and normalized, to respect the energy
conservation principle:~
\begin{equation}
w_{\vec{o}_{k}j}^{*}\in\mathbb{R^{+}},\,\sum_{\forall\vec{o}}w_{\vec{o}_{k}j}^{*}\le1
\end{equation}
.}
\item \textcolor{black}{The values sourced from a node (downstream) are
fully defined by the node's virtual norm $\hat{n}_{j}$, i.e., combining
eq.~(\ref{eq:lawReflect}) with (\ref{eq:weights-1}) (omitting the
$\max$ operator and the normalization for ease of exposition):~$w_{\vec{o}_{k}j}^{*}=\vec{o_{k}}\cdot\sum_{\forall i}\left(\vec{d_{ji}}-2\left(\vec{d_{ji}}\cdot\hat{n_{j}}\right)\cdot\hat{n}_{j}\right),$reminding
that $\vec{o_{k}}$ and $\vec{d_{ji}}$ are static and constant vectors,
defined by the floorplan geometry ant SDM tile locations. }
\end{itemize}
\textcolor{black}{The virtual normal $\hat{n_{j}}$ is furthermore
defined completely by any two of the rotation matrix angles $\left\langle \theta_{j},\phi_{j},\varphi_{j}\right\rangle $,
as shown in eq.~(\ref{eq:weights}). For ease of exposition we will
keep the pair $\left\langle \theta_{j},\phi_{j}\right\rangle $. Then,
we can write that $w_{\vec{o}_{k}j}^{*}=\mathcal{F}_{\vec{o}_{k}j}\left(\theta_{j},\phi_{j}\right)$
with $\mathcal{F}$ denoting the relation specialized for each $\vec{o}_{k}j$
index. In other words, the weights are not free to take any values
and are defined by $\left\langle \theta_{j},\phi_{j}\right\rangle $.
As such, the update rule~(\ref{eq:updateGeneric}) needs to be updated
accordingly as follows. Our entry-point will be the $\eta$ constant,
which will be specialized for each $ij$ index as $\eta_{ij}$. We
note that this practice is not uncommon: weight freezing approaches
make use of this technique~\cite{islam2001new}. Finally, we set
$\eta_{ij}$ as:}\textcolor{black}{\small{}
\begin{multline}
\eta_{ij}\gets\arg opt_{\eta_{ij}}\left(w_{ji}-\eta_{ij}\cdot\frac{\partial\xi}{\partial w_{ji}}\right),\\
\text{\textbf{subject\,to}:}\\
\exists\left\langle \theta_{j},\phi_{j}\right\rangle :w_{ji}-\eta_{ij}\cdot\frac{\partial\xi}{\partial w_{ji}}=\mathcal{F}_{\vec{o}_{k}j}\left(\theta_{j},\phi_{j}\right),\forall j\\
\text{\ensuremath{\left(w_{ji}-\eta_{ij}\cdot\frac{\partial\xi}{\partial w_{ji}}\right)\in}\ensuremath{\mathbb{R^{+}}},\ensuremath{\sum_{\forall\vec{o}}\left(w_{ji}-\eta_{ij}\cdot\frac{\partial\xi}{\partial w_{ji}}\right)\le}1}\label{eq:weightalter}
\end{multline}
}\textcolor{black}{or in an approximate form as:}\textcolor{black}{\small{}
\begin{multline}
\eta_{ij}\gets\arg\min_{\eta_{ij}}\left(\sum_{\forall j}\left|\left(w_{ji}-\eta_{ij}\cdot\frac{\partial\xi}{\partial w_{ji}}\right)-\left(\mathcal{F}_{\vec{o}_{k}j}\left(\theta_{j},\phi_{j}\right)\right)\right|\right),\\
\text{\textbf{subject\,to:}}\\
\left(w_{ji}-\eta_{ij}\cdot\frac{\partial\xi}{\partial w_{ji}}\right)\in\mathbb{R^{+}},\,\sum_{\forall\vec{o}}\left(w_{ji}-\eta_{ij}\cdot\frac{\partial\xi}{\partial w_{ji}}\right)\le1\,\,\,\blacksquare
\end{multline}
}{\small\par}
\end{document}